\documentclass[12pt,preprint]{aastex}
\begin{document}

\title{Relativistic Equation of State for Core-Collapse Supernova Simulations}

\author{H. Shen}
\affil{School of Physics, Nankai University, Tianjin 300071, China}
\email{shennankai@gmail.com}

\author{H. Toki}
\affil{Research Center for Nuclear Physics (RCNP), Osaka University,
       Ibaraki, Osaka 567-0047, Japan}
\email{toki@rcnp.osaka-u.ac.jp}

\author{K. Oyamatsu}
\affil{Department of Human Informatics, Aichi Shukutoku University,
       Nagakute-cho, Aichi 480-1197, Japan}
\email{oyak@asu.aasa.ac.jp}

\and

\author{K. Sumiyoshi}
\affil{Numazu College of Technology, Ooka 3600, Numazu, Shizuoka 410-8501, Japan; \\
       Theory Center, High Energy Accelerator Research Organization (KEK),
       Oho 1-1, Tsukuba 305-0801, Japan}
\email{sumi@numazu-ct.ac.jp}

%%%%%%%%%%%%%%%%%%%%%%%%%%%%%%%%%%%%%%%%%%%%%%%%%%%%%%%%%%%%%%%%%%%%%%%%%%%%%%%%
\begin{abstract}
We construct the equation of state (EOS) of dense matter covering a wide
range of temperature, proton fraction, and density for the use of core-collapse
supernova simulations. The study is based on the relativistic mean-field (RMF)
theory, which can provide an excellent description of nuclear matter and
finite nuclei. The Thomas--Fermi approximation in combination with
assumed nucleon distribution functions and a free energy minimization
is adopted to describe the
non-uniform matter, which is composed of a lattice of heavy nuclei.
We treat the uniform matter and non-uniform matter consistently using
the same RMF theory. We present two sets of EOS tables, namely EOS2 and EOS3.
EOS2 is an update of our earlier work published in 1998 (EOS1),
where only the nucleon degree of freedom is taken into account.
EOS3 includes additional contributions from $\Lambda$ hyperons.
The effect of $\Lambda$ hyperons on the EOS is negligible in the low-temperature
and low-density region, whereas it tends to soften the EOS at high density.
In comparison with EOS1, EOS2 and EOS3 have an improved design of
ranges and grids, which covers the temperature range $T=0.1$--$10^{2.6}$ MeV
with the logarithmic grid spacing $\Delta \log_{10}(T/\rm{[MeV]})=0.04$
(92 points including $T=0$), the proton fraction range $Y_p=0$--$0.65$
with the linear grid spacing $\Delta Y_p = 0.01$ (66 points),
and the density range $\rho_B=10^{5.1}$--$10^{16}\,\rm{g\,cm^{-3}}$
with the logarithmic grid spacing
$\Delta \log_{10}(\rho_B/\rm{[g\,cm^{-3}]}) = 0.1$ (110 points).
\end{abstract}


%%%%%%%%%%%%%%%%%%%%%%%%%%%%%%%%%%%%%%%%%%%%%%%%%%%%%%%%%%%%%%%%%%%%%%%%%%%%%%%%
\section{Introduction}
\label{sec:1}

The equation of state (EOS) of dense matter plays an important role in various
astrophysical phenomena such as supernova explosions and the formation of
neutron stars and black holes~\citep{jank07,burr06,sumi05,sumi09}.
Simulations of core-collapse supernovae cover a wide range of
thermodynamic conditions, and extremely high density and temperature
may be achieved when black holes are formed by failed supernovae.
The temperature may vary from $0$ to more than $100$ MeV,
the proton fraction changes from $0$ to around $0.6$,
and the density can vary from $10^{5}$ to more than $10^{15}\,\rm{g\,cm^{-3}}$.
Clearly, it is very difficult to construct a
complete EOS over such a wide range of parameters.
The information of matter under extreme conditions is far beyond our knowledge
of nuclear physics from laboratory experiments. Therefore, it is necessary to
perform a large extrapolation based on a theoretical model that is
supported by microscopic theory and consistent with available
experimental data. During the past few decades, great efforts have been made
to study the EOS of nuclear matter~\citep{latt91,scha96,webe05,latt07}.
However, most of the investigations focused on detailed aspects of nuclear
matter, which were often restricted to the case of zero temperature or high
density with uniform distribution of particles. This kind of EOS is generally
not applicable for use in supernova simulations.
So far, there exist only two realistic EOSs which are commonly used in
simulations of core-collapse supernovae,
namely the one by~\citet{latt91} and the one by~\citet{shen98b}.
The Lattimer--Swesty EOS is based on a compressible liquid-drop model
with a Skyrme force. The Shen EOS is based on a relativistic mean-field
(RMF) model and uses the Thomas--Fermi approximation with assumed nucleon
distribution functions in a Wigner--Seitz
cell for the description of non-uniform matter.
Recently, a Hartree mean-field calculation~\citep{shen10a} was performed for the
Wigner--Seitz cell instead of the Thomas--Fermi approximation used
by~\citet{shen98b}. The Hartree calculation can incorporate nuclear shell
effects, but it requires much more computational resources.

In our earlier work~\citep{shen98a,shen98b}, we constructed the relativistic
EOS for supernova simulations, which has been widely used in astrophysical
simulations over the past decade~\citep{jank07,burr06,sumi05,sage09}.
The EOS is based on the RMF theory combined with the Thomas--Fermi
approximation. The RMF theory with nonlinear $\sigma$ and $\omega$
terms is able to reproduce nuclear matter saturation properties
and provide a good description for both stable and unstable nuclei~\citep{suga94a,hira96}.
The Thomas--Fermi approximation in combination with assumed nucleon distribution functions
and a free energy minimization is adopted to describe the non-uniform
matter which is modeled as a mixture of a single species of heavy
nuclei, alpha particles, and free nucleons that exist outside of nuclei,
while the leptons can be treated as uniform non-interacting particles
separately. The RMF results are taken as input in the Thomas--Fermi calculation,
so the treatments of non-uniform matter and uniform
matter in this EOS are sufficient to obtain the table in a consistent manner.
It would be preferable to treat the mixture of nuclei as recently done
in~\citet{hemp10},~\citet{furu11}, and~\citet{blin11} for the detailed treatment of
electron captures on nuclei in supernova core.
However, it is beyond the scope of the current update of the Shen EOS table,
which has been routinely used in astrophysical simulations.

Toward a more efficient and effective use of the Shen EOS table,
we are required to make improvements in the EOS given in~\citet{shen98b},
hereafter referred to as EOS1.
The main demand is to increase the number of temperature points
which is crucial in the simulation of core-collapse supernovae.
In some cases, the temperature may reach more than $100$ MeV,
so it is encouraged to provide results at some higher temperatures
although it is probably beyond the applicability of the RMF theory.
It is generally believed that at sufficiently high temperature and/or
density nuclear matter undergoes a phase transition to quark--gluon
plasma (QGP). Based on experimental data from high-energy heavy ion
collisions and lattice quantum chromodynamics (QCD) calculations, it is known that the critical
temperature for the QCD phase transition is around $T_c=175$ MeV
for zero baryon density~\citep{gupt11}. Therefore, we note that
the nuclear EOS at extremely high temperature and/or density is not
reliable due to the QCD phase transition, although results of the RMF
theory are provided for practical use in astrophysical simulations.
Another suggestion is to use a linear grid for the proton fraction $Y_p$,
instead of the logarithmic $Y_p$ grid used in EOS1, which can add
more points in the important region $Y_p\sim 0.2$--$0.5$ for supernova
simulations and save memory space by reducing the number of $Y_p$
points at $Y_p<0.1$.
Furthermore, a proton-rich matter may be involved in astrophysical
simulations~\citep{prue05,froh06},
and therefore the information at $Y_p\sim 0.6$ is needed.
In EOS1, the density grid spacing is only approximately equal at
high density, it is now possible to perform the calculation
with equal grid spacing in the whole range of density.
Since it is desirable and feasible to make these improvements,
we work out a new version of the EOS table,
hereafter referred to as EOS2, which contains the same degrees
of freedom as EOS1.

In recent years, there has been extensive discussion in the literature
on the influence of non-nucleonic degrees of freedom in dense
matter~\citep{webe05,latt07}.
It is generally believed that hyperons appear around twice normal
nuclear matter density in cold neutron star matter~\citep{scha96,shen02,ishi08}.
The first hyperon to appear is $\Lambda$ that is the lightest one
with an attractive potential in nuclear matter~\citep{ishi08,yuep09}.
$\Sigma$ hyperons are now considered to appear at a higher density than
$\Lambda$, because $\Sigma$ hyperons feel a repulsive potential in
nuclear matter according to recent developments in hypernuclear physics~\citep{ishi08}.
In the work of~\citet{ishi08}, the authors examined the properties of dense matter
based on an extended RMF model including the full baryon octet.
They presented several sets of EOS including hyperons for simulations
of core-collapse supernovae, which were connected with EOS1 at low density
in the simple procedure described in Section 2.3 of their paper.
The effect of $s$-wave pion condensation was also examined in~\citet{ishi08},
where the authors considered free thermal pions assuming the pion mass could
not be affected by the interaction. In~\citet{ohni09}, the authors examined
the possibility of $s$-wave pion condensation in dense matter by using the
phenomenological optical potentials determined from the pionic atom or
pion--nucleus scattering data, and they concluded that $s$-wave pion condensation
would hardly take place in neutron stars and especially have no chance
if hyperons could participate in neutron star matter.
The presence of boson condensation and deconfined quarks in neutron stars
has been extensively discussed in many works~\citep{scha96,webe05,latt07,yang08}.
It has been suggested that the quark matter may exist in the core of
massive neutron stars, and the hadron--quark phase transition can
proceed through a mixed phase of hadronic and quark
matter~\citep{glen92,webe05,latt07,yang08}. If deconfined quark matter does
exist inside stars, it is likely to be in a color superconducting
phase, and various color superconducting phases
have been intensively investigated in recent years~\citep{buba05,paul11}.
In the works of~\citet{naka08,naka10} and~\citet{sage09,sage10},
the authors constructed the EOS tables for simulations of core-collapse
supernovae including the hadron--quark phase transition at high density,
which were connected with EOS1 at low density.
The hadron--quark phase transition proceeded through a mixed phase obtained
by the Gibbs conditions for phase equilibrium, where the RMF model was used
for the hadronic phase and the bag model was adopted for the quark phase.
Generally, the introduction of non-nucleonic degrees of freedom leads to
a softening of the EOS and thereby a corresponding reduction in the maximum
mass of neutron stars.
The recent measurement of the Shapiro delay in the radio pulsar PSR J1614-2230
yielded a mass of $1.97\pm 0.04\, M_{\odot}$~\citep{demo10}.
Such a high neutron star mass provides an important constraint on the EOS
at high density and rules out many predictions of non-nucleonic components
in neutron star interiors.
However, it is currently difficult to rule out all possible exotica
with the $1.97\, M_{\odot}$ observation, some theoretical calculations
including hyperons and/or quarks could be compatible with the observation
of PSR J1614-2230~\citep{latt07,latt10,ston07,paul11}.

Among these exotic candidates, $\Lambda$ hyperons are the most likely
to occur in dense matter.
In addition, much more experimental information
is now available for $\Lambda$ than other hyperons~\citep{taka01,nakazawa10}.
From the experimental binding energies of single-$\Lambda$ hypernuclei,
the potential depth of $\Lambda$ in nuclear matter is estimated to be
around $-30$ MeV~\citep{shen06}.
Several recent observations of double-$\Lambda$ hypernuclei (see Table 4 of~\citet{nakazawa10})
indicate that the effective $\Lambda\Lambda$ interaction should be considerably
weaker than that deduced from the earlier measurement~\citep{hiya10}.
In theoretical studies of $\Lambda$ hypernuclei, the RMF theory with
nonlinear $\sigma$ and $\omega$ terms is able to provide a reasonable
description of single- and double-$\Lambda$ hypernuclei~\citep{shen06}.
The influence of $\Lambda$ hyperons on neutron star properties has
been investigated within the RMF model~\citep{suga94b}.
We would like to examine the effect of $\Lambda$ hyperons on the EOS
for simulations of core-collapse supernovae, while other hyperons
such as $\Sigma$ and $\Xi$ are ignored due to their relatively
high threshold densities and lack of available experimental data.
We construct the relativistic EOS with the inclusion of $\Lambda$
hyperons, hereafter referred to as EOS3.
For the contribution of $\Lambda$ hyperons in EOS3, we assume that
the equilibrium condition $\mu_n=\mu_{\Lambda}$ is valid.
In a supernova explosion, the dynamical timescale is of the order of
milliseconds, which is long enough to establish equilibrium with
respect to weak interactions that change strangeness on the timescale
of microseconds or less~\citep{sage09,sage10}.
In fact, $\Lambda$ hyperons can have a noticeable effect on
the properties of matter mainly at high density.
It is known that weak equilibrium could be achieved at densities
above $10^{12}\,\rm{g\,cm^{-3}}$ where neutrinos begin to be trapped
in the core~\citep{kota06,jank07}. Therefore, it is justified to
use the equilibrium condition $\mu_n=\mu_{\Lambda}$
to determine the $\Lambda$ fraction of supernova matter at high density
where $\Lambda$ hyperons have a noticeable contribution.
In the work of~\citet{ishi08}, the authors assumed weak equilibrium
among the full baryon octet when they determined the composition
of supernova matter at a fixed density, temperature, and charge fraction.
They examined the appearance of hyperons during the evolution of core
collapse and bounce, and found that the effect of hyperons would be
small in a spherical and adiabatic collapse of a $15\, M_{\odot}$ star
by the hydrodynamics without neutrino transfer~\citep{ishi08}.
The hyperons are expected to appear and play an important role in the thermal
evolution of protoneutron stars and the black hole formation from massive
stars~\citep{pons99,sumi09}.

We have two aims in this paper. The first is to improve the EOS table
according to the requirements of the users.
In comparison with the earlier version (EOS1), the following improvements
are made in EOS2 and EOS3 (see Table 1 for details of the comparison).
\begin{itemize}
\item
 The number of $T$ points is largely increased,
 the upper limit of $T$ is extended,
 and equal grid spacing for $T$ is used.
\item
 A linear $Y_p$ grid is adopted instead of the logarithmic $Y_p$ grid used in EOS1
 and the $Y_p$ upper limit is extended.
\item
 The upper limit of $\rho_B$ is extended and equal grid spacing
 is used in the whole range of $\rho_B$,
 whereas it is only approximately equal at high density in EOS1.
\end{itemize}
The finer grids are favorable for better accuracy in the numerical simulations of
core-collapse supernovae. The wide coverage of the conditions is necessary for
numerical simulations of astrophysical phenomena including black hole formation,
neutron star mergers, and nucleosynthesis.
The second aim of this paper is to provide the EOS table with the
inclusion of $\Lambda$ hyperons for the use in simulations of
core-collapse supernovae. The difference between EOS2 and EOS3
is that the contribution from $\Lambda$ hyperons is included in EOS3
when the $\Lambda$ fraction is larger than $10^{-5}$.

This paper is arranged as follows. In Section~\ref{sec:2}, we briefly
describe the framework to calculate the EOS table. We introduce the
RMF model with the inclusion of $\Lambda$ hyperons and explain how
to determine the RMF parameters. To make this paper self-contained,
we give a brief description of the Thomas--Fermi approximation for
the non-uniform matter.
In Section~\ref{sec:3}, we discuss our results without and with
the inclusion of $\Lambda$ hyperons. We explore the properties of
dense matter by examining the phase diagram, compositions, and
thermodynamic quantities.
Section~\ref{sec:4} is devoted to a summary.
In Appendix A, we give the definitions of the physical quantities tabulated in the EOS.
EOS2 and EOS3 are presented in electronic tables and can also be found on several Web sites.
In Appendix B, we describe various checks made for the EOS tables.

%%%%%%%%%%%%%%%%%%%%%%%%%%%%%%%%%%%%%%%%%%%%%%%%%%%%%%%%%%%%%%%%%%%%%%%%%%%%%%%%
\section{Model descriptions}
\label{sec:2}

We construct the EOS within the RMF framework~\citep{sero86,shen98a,shen98b,shen06}.
For uniform matter without heavy nuclei formed, the RMF theory can be easily used
to calculate the properties of matter. For non-uniform matter where heavy nuclei
are formed in order to lower the free energy, we adopt the Thomas--Fermi
approximation based on the work by~\citet{oyam93}.
The non-uniform matter can be modeled as a mixture of a single species of
heavy nuclei, alpha particles, and free nucleons that exist outside of nuclei.
The results of the RMF model are taken as input in the Thomas--Fermi calculation,
so the treatments of non-uniform matter and uniform matter
in this EOS are sufficient to obtain the table in a consistent manner.
It would be preferable to treat the mixture of nuclei as recently done
in~\citet{hemp10},~\citet{furu11}, and~\citet{blin11} for the detailed treatment of
electron captures on nuclei in supernova core.
However, it is beyond the scope of the current update of the Shen EOS table,
which has been routinely used in astrophysical simulations.

%%%%%%%%%%%%%%%%%%%%%%%%%%%%%%%%%%%%%%%%
\subsection{Relativistic mean-field theory}
\label{sec:2.1}

We adopt the RMF theory with nonlinear $\sigma$ and $\omega$ terms
to describe homogeneous nuclear matter~\citep{suga94a,shen06}.
We study the two cases without and with $\Lambda$ hyperons.
In the RMF approach, baryons interact through
the exchange of various effective mesons.
The exchanged mesons considered in this work include isoscalar scalar and
vector mesons ($\sigma$ and $\omega$) and an isovector vector meson ($\rho$).
In some published studies~\citep{scha96,shen06,yuep09}, the two hidden-strangeness
($ \bar{s} s $) scalar and vector mesons
($\sigma^{\ast}$ and $\phi$) were included in a hyperon-rich system.
It has been found that the attraction from $\sigma^{\ast}$ exchange is almost
canceled by the repulsion from $\phi$ exchange~\citep{shen06}.
Therefore, we neglect the contribution from the exchange of these two
hidden-strangeness mesons in this work.

We start with the Lagrangian of the RMF theory including $\Lambda$ hyperons,
\begin{eqnarray}
\label{eq:LRMF}
{\cal L}_{\rm{RMF}} & = & \bar{\psi}\left[i\gamma_{\mu}\partial^{\mu} -M
-g_{\sigma}\sigma-g_{\omega}\gamma_{\mu}\omega^{\mu}
-g_{\rho}\gamma_{\mu}\tau_a\rho^{a\mu}
\right]\psi  \nonumber\\
& & +\bar{\psi}_{\Lambda}\left[i\gamma_{\mu}\partial^{\mu} -M_{\Lambda}
-g_{\sigma}^{\Lambda}\sigma-g_{\omega}^{\Lambda}\gamma_{\mu}\omega^{\mu}
\right]\psi_{\Lambda}  \nonumber\\
 && +\frac{1}{2}\partial_{\mu}\sigma\partial^{\mu}\sigma
-\frac{1}{2}m^2_{\sigma}\sigma^2-\frac{1}{3}g_{2}\sigma^{3}
-\frac{1}{4}g_{3}\sigma^{4} \nonumber\\
 && -\frac{1}{4}W_{\mu\nu}W^{\mu\nu}
+\frac{1}{2}m^2_{\omega}\omega_{\mu}\omega^{\mu}
+\frac{1}{4}c_{3}\left(\omega_{\mu}\omega^{\mu}\right)^2   \nonumber\\
 && -\frac{1}{4}R^a_{\mu\nu}R^{a\mu\nu}
+\frac{1}{2}m^2_{\rho}\rho^a_{\mu}\rho^{a\mu} ,
\end{eqnarray}
where $\psi$ and $\psi_{\Lambda}$ denote the nucleon and
$\Lambda$ hyperon fields, respectively.
$\sigma$, $\omega^{\mu}$, and $\rho^{a\mu}$ are
$\sigma$, $\omega$, and $\rho$ meson fields with masses
$m_{\sigma}$, $m_{\omega}$, and $m_{\rho}$.
$W^{\mu\nu}$ and $R^{a\mu\nu}$ are the antisymmetric field tensors
for $\omega^{\mu}$ and  $\rho^{a\mu}$, respectively.
It is known that the inclusion of nonlinear $\sigma$ terms is
essential to reproduce the properties of nuclei quantitatively and provide
a reasonable value for the incompressibility, while the nonlinear $\omega$
term is added in order to reproduce the density dependence of the vector part
of the nucleon self-energy obtained
in the relativistic Brueckner Hartree--Fock (RBHF) theory~\citep{suga94a}.
We adopt the parameter set TM1 listed in Table~\ref{tab:2},
which was determined in~\citet{suga94a} by fitting some ground-state
properties of nuclei including unstable ones.
With the TM1 parameter set, the nuclear matter saturation
density is $0.145$ fm$^{-3}$, the binding energy per nucleon is $16.3$ MeV,
the symmetry energy is $36.9$ MeV, and the incompressibility is $281$ MeV.
The RMF theory with the TM1 parameter set provides an excellent description
of nuclear matter and finite nuclei~\citep{suga94a},
and it is also shown to agree satisfactorily with experimental data
in studies of nuclei with deformed configurations~\citep{hira96}.
For the parameters of $\Lambda$ hyperons, we use the experimental mass value
$M_{\Lambda}=1115.7$ MeV~\citep{amsl08}. Concerning the coupling constants between mesons
and $\Lambda$ hyperons, we take $g_{\omega}^{\Lambda}/g_{\omega}=2/3$ based
on the naive quark model and $g_{\sigma}^{\Lambda}/g_{\sigma}=0.621$ determined
by fitting experimental binding energies of single-$\Lambda$ hypernuclei~\citep{shen06},
which produce an attractive potential of $\Lambda$ in nuclear matter
at saturation density to be around $-30$ MeV.
It is known that the inclusion of tensor coupling between $\omega$ and $\Lambda$ is
important to produce small spin-orbit splittings of single-$\Lambda$ hypernuclei,
but it does not contribute to homogeneous matter.
The $\Lambda$ hyperon is a charge neutral and isoscalar particle,
so that it does not couple to the $\rho$ meson.
It is shown that these parameters can reproduce well the experimental data
for both single- and double-$\Lambda$ hypernuclei~\citep{shen06}.

Starting with the Lagrangian~(\ref{eq:LRMF}), we derive a set of Euler--Lagrange equations.
We employ the RMF approximation as described in~\citet{sero86},
where the meson fields are treated as classical fields
and the field operators are replaced by their expectation values.
For homogeneous matter, the non-vanishing expectation values of meson fields
are   $\sigma =\left\langle \sigma    \right\rangle$,
      $\omega =\left\langle \omega^{0}\right\rangle$,
and   $\rho   =\left\langle \rho^{30} \right\rangle$.
The equations of motion for the meson fields in homogeneous matter
have the following form:
\begin{eqnarray}
\label{eq:ms0}
\sigma & = &
 -\frac{g_{\sigma}}{m_{\sigma}^2} \langle\bar{\psi}\psi\rangle
 -\frac{g_{\sigma}^{\Lambda}}{m_{\sigma}^2} \langle\bar{\psi_{\Lambda}}\psi_{\Lambda}\rangle
 -\frac{1}{m_{\sigma}^2}\left(g_{2}\sigma^{2}+g_{3}\sigma^{3}\right),
\\
\label{eq:mw0}
\omega & = &
  \frac{g_{\omega}}{m_{\omega}^2} \langle\bar{\psi}\gamma^0\psi\rangle
 +\frac{g_{\omega}^{\Lambda}}{m_{\omega}^2} \langle\bar{\psi}_{\Lambda}\gamma^0\psi_{\Lambda}\rangle
 -\frac{1}{m_{\omega}^2}c_3\omega^3,
\\
\label{eq:mr0}
\rho   & = &
 \frac{g_{\rho}}{m_{\rho}^2} \langle\bar{\psi}\tau_3\gamma^0\psi\rangle.
\end{eqnarray}
The stationary Dirac equations for nucleons and $\Lambda$ hyperons are given by
\begin{eqnarray}
\label{eq:driacn}
 & & \left(-i\alpha_k\nabla^k + \beta M^{*}_N
+g_{\omega}\omega +g_{\rho}\tau_3\rho \right)\psi_{N}^{s}=
\varepsilon_{N}^{s} \psi_{N}^{s} ,
\\
\label{eq:driacl}
 & & \left(-i\alpha_k\nabla^k + \beta M^{*}_{\Lambda}
+g_{\omega}^{\Lambda}\omega \right)\psi_{\Lambda}^{s}=
\varepsilon_{\Lambda}^{s} \psi_{\Lambda}^{s},
\end{eqnarray}
where $N$ stands for the nucleons ($N=p$ or $n$).
$M^{*}_N=M+g_{\sigma}\sigma$ and
$M^{*}_{\Lambda}=M_{\Lambda}+g_{\sigma}^{\Lambda}\sigma$
are the effective nucleon mass and effective $\Lambda$ mass, respectively.
$s$ denotes the index of eigenstates,
while $\varepsilon_{N}^{s}$ and $\varepsilon_{\Lambda}^{s}$
are the single-particle energies.

In homogeneous matter, baryons occupy single-particle states with
the occupation probability $f_{i}^{s}$ ($i=p$, $n$, or $\Lambda$).
At zero temperature, $f_{i}^{s}=1$ under the Fermi surface,
while $f_{i}^{s}=0$ above the Fermi surface.
For finite temperature, the occupation probability is given by the Fermi--Dirac
distribution,
\begin{eqnarray}
 f_{i}^{s}=\frac{1}{1+\exp\left[\left(\varepsilon_{i}^{s}-\mu_{i}\right)/T\right]}
       =\frac{1}{1+\exp\left[\left(\sqrt{k^2+{M_{i}^{*}}^2}-\nu_{i}\right)
        /T\right]},
\\
 f_{\bar{i}}^{s}=\frac{1}{1+\exp\left[\left(-\varepsilon_{\bar{i}}^{s}+\mu_{i}
       \right)/T\right]}
 =\frac{1}{1+\exp\left[\left(\sqrt{k^2+{M_{i}^{*}}^2}+\nu_{i}\right)/T\right]} ,
\end{eqnarray}
where $i$ and $\bar{i}$ denote the particle and antiparticle, respectively.
$\varepsilon_{i}^{s}$ and $\varepsilon_{\bar{i}}^{s}$ are the single-particle energies.
The relation between the chemical potential $\mu_i$ and the kinetic
part of the chemical potential $\nu_i$ is given by
\begin{eqnarray}
 \mu_{p}       &=& \nu_p      +g_{\omega}\omega +g_{\rho}\rho, \\
 \mu_{n}       &=& \nu_n      +g_{\omega}\omega -g_{\rho}\rho, \\
 \mu_{\Lambda} &=& \nu_\Lambda+g_{\omega}^{\Lambda}\omega.
\end{eqnarray}
The chemical potential $\mu_i$ is related to the baryon number density
$n_i$ as
\begin{equation}
\label{eq:nirmf}
 n_{i}=\frac{1}{\pi^2}
       \int_0^{\infty} dk\,k^2\,\left(f_{i}^{k}-f_{\bar{i}}^{k}\right),
\end{equation}
where the quantum number $s$ is replaced by the momentum $k$ when we do the
integration in the momentum space instead of summing over the eigenstates.
We denote $n_B=n_p+n_n+n_{\Lambda}$ as the total baryon number density
and $Y_p=n_p / n_B $ as the proton fraction in homogeneous matter.
The coupled equations are solved at fixed $n_B$ and $Y_p$.
At zero temperature, $\Lambda$ hyperons appear only at high density when
the equilibrium condition $\mu_n=\mu_{\Lambda}$ could be satisfied.
For finite temperature, a small number of $\Lambda$ hyperons may exist
at low density with $\mu_n=\mu_{\Lambda}$, and the $\Lambda$ fraction
increases rapidly at high density.
The thermodynamic quantities of homogeneous matter have been derived
in~\citet{sero86},~\citet{shen98a,shen98b}, and~\citet{shen02}, so we simply write the expressions here.
The energy density of nuclear matter including $\Lambda$ hyperons is given by
\begin{eqnarray}
\epsilon &=& \displaystyle{\sum_{i=p,n,\Lambda} \frac{1}{\pi^2}
   \int_0^{\infty} dk\,k^2\,
   \sqrt{k^2+{M_{i}^*}^2}  \left(f_{i}^{k}+f_{\bar{i}}^{k}\right) } \nonumber\\
 & &
  +\frac{1}{2}m_{\sigma}^2\sigma^2+\frac{1}{3}g_{2}\sigma^{3}+\frac{1}{4}g_{3}\sigma^{4}
  +\frac{1}{2}m_{\omega}^2\omega^2+\frac{3}{4}c_{3}\omega^{4}
  +\frac{1}{2}m_{\rho}^2\rho^2,
\end{eqnarray}
the pressure is given by
\begin{eqnarray}
 p &=& \displaystyle{\sum_{i=p,n,\Lambda} \frac{1}{3\pi^2} \int_0^{\infty} dk\,k^2\,
   \frac{k^2}{\sqrt{k^2+{M_{i}^*}^2}}  \left(f_{i}^{k}+f_{\bar{i}}^{k}\right) } \nonumber\\
 & &
  -\frac{1}{2}m_{\sigma}^2\sigma^2-\frac{1}{3}g_{2}\sigma^{3}-\frac{1}{4}g_{3}\sigma^{4}
  +\frac{1}{2}m_{\omega}^2\omega^2+\frac{1}{4}c_{3}\omega^{4}
  +\frac{1}{2}m_{\rho}^2\rho^2,
\end{eqnarray}
and the entropy density is given by
\begin{eqnarray}
s & =  \displaystyle{\sum_{i=p,n,\Lambda} \frac{1}{\pi^2} \int_0^{\infty} dk\,k^2 }
   & \left[ -f_{i}^{k}\ln f_{i}^{k}
            -\left(1-f_{i}^{k}\right)\ln \left(1-f_{i}^{k}\right) \right. \nonumber\\
 & & \left. -f_{\bar{i}}^{k}\ln f_{\bar{i}}^{k}
            -\left(1-f_{\bar{i}}^{k}\right)\ln \left(1-f_{\bar{i}}^{k}\right) \right] .
\end{eqnarray}

%%%%%%%%%%%%%%%%%%%%%%%%%%%%%%%%%%%%%%%%
\subsection{Thomas--Fermi approximation}
\label{sec:2.2}

In the low-temperature and low-density region, heavy nuclei may be
formed in order to lower the free energy.
We adopt the Thomas--Fermi approximation in combination with assumed
nucleon distribution functions and a free energy minimization
to describe the non-uniform matter based on the work by~\citet{oyam93}.
In this study, we take into account the contribution from $\Lambda$ hyperons
only when the $\Lambda$ fraction $X_{\Lambda}$ is larger than $10^{-5}$.
In the region where the heavy nuclei are formed, $X_{\Lambda}$ is quite small,
therefore we neglect the contribution from $\Lambda$ hyperons
in the Thomas--Fermi calculation.
The non-uniform matter is modeled as a mixture of a single species of
heavy nuclei, alpha particles, and free nucleons that exist outside of nuclei,
while the leptons can be treated as uniform non-interacting particles separately.
For the system with a fixed proton fraction, the leptons play no role in the
free energy minimization. Hence we mainly pay attention to the baryon
contribution in this study.

We assume that each heavy spherical nucleus is
located in the center of a charge-neutral cell consisting of a
vapor of neutrons, protons, and alpha-particles. The nuclei form a
body-centered-cubic (BCC) lattice to minimize the Coulomb lattice energy.
It is useful to introduce the Wigner--Seitz cell to simplify the energy of
a unit cell. The Wigner--Seitz cell is a sphere whose volume is
the same as the unit cell in the BCC lattice.
The lattice constant $a$ is defined as the cube root of the cell volume,
$ V_{\rm{cell}}=a^3=N_B / n_B $,
where $N_B$ and $n_{B}$ are the
baryon number per cell and the average baryon number density, respectively.
We define the baryon mass density as $\rho_B=m_{u} n_B$ with
$m_{u}$ being the atomic mass unit~\citep{amsl08}.
We calculate the Coulomb energy using the Wigner--Seitz approximation and
adding an energy correction for the BCC lattice~\citep{oyam93}.
This energy correction is negligible unless the nuclear size is comparable to the cell size.

We assume the nucleon distribution function $n_i(r)$ ($i=p$ or $n$)
in the Wigner--Seitz cell as
\begin{equation}
\label{eq:nitf}
n_i\left(r\right)=\left\{
\begin{array}{ll}
\left(n_i^{\rm{in}}-n_i^{\rm{out}}\right) \left[1-\left(\frac{r}{R_i}\right)^{t_i}
\right]^3 +n_i^{\rm{out}},  & 0 \leq r \leq R_i, \\
n_i^{\rm{out}},  & R_i \leq r \leq R_{\rm{cell}}, \\
\end{array} \right.
\end{equation}
where $r$ represents the distance from the center of the nucleus
and $R_{\rm{cell}}$ is the radius of the Wigner--Seitz cell defined by
$ V_{\rm{cell}} = 4 \pi R_{\rm{cell}}^3 / 3 $.
The density parameters
$n_i^{\rm{in}}$ and $n_i^{\rm{out}}$ are
the densities at $r=0$ and $ r \geq R_i $, respectively. The parameters
$R_i$ and $t_i$ determine the boundary and the relative surface thickness
of the nucleus.
For the distribution function of alpha-particle $n_{\alpha}(r)$,
which should decrease as $r$ approaches the center of the nucleus,
we assume
\begin{equation}
\label{eq:na}
n_{\alpha}\left(r\right)=\left\{
\begin{array}{ll}
-n_{\alpha}^{\rm{out}} \left[1-\left(\frac{r}{R_p}\right)^{t_p}
\right]^3 +n_{\alpha}^{\rm{out}},  & 0 \leq r \leq R_p, \\
n_{\alpha}^{\rm{out}},  & R_{p} \leq r \leq R_{\rm{cell}}, \\
\end{array} \right.
\end{equation}
which could give $n_{\alpha} (r=0) =0$
and $n_{\alpha} (r>R_p) =n_{\alpha}^{\rm{out}}$.
Here we use the same parameters $R_p$ and $t_p$ for both proton and
alpha-particle distribution functions in order to avoid the presence of
too many parameters in the minimization procedure.
The parameters $R_n$ and $t_n$ may be somewhat different from
$R_p$ and $t_p$ due to the additional neutrons forming a neutron skin
in the surface region.
For a system with fixed temperature $T$, proton fraction $Y_p$,
and baryon mass density $\rho_B$, there are eight independent parameters
among the ten variables, $a$, $n_n^{\rm{in}}$, $n_n^{\rm{out}}$, $R_n$,
$t_n$, $n_p^{\rm{in}}$, $n_p^{\rm{out}}$, $R_p$, $t_p$, and $n_{\alpha}^{\rm{out}}$.
The thermodynamically favorable state is the one that minimizes
the free energy density with respect to these eight independent parameters.
In principle, the resulting nucleon distribution in the Wigner--Seitz cell
would depend on the form of the parameterization.
It is also possible to determine the nucleon distribution by a self-consistent
Thomas--Fermi method without any form of parameterization~\citep{sil01,shen10b}.
We have compared, in Figures 1 and 2 of~\citet{shen10b}, the results obtained
by the self-consistent Thomas--Fermi method with those using the parameterization
of Equation~(\ref{eq:nitf}), and found that they were in good agreement with each
other for the cases considered.
Therefore, Equation~(\ref{eq:nitf}) is considered to be a reasonable
form of the nucleon distribution in the Wigner--Seitz cell.

In this model, the free energy density contributed from baryons is given by
\begin{equation}
f\,=\,F_{\rm{cell}}\,/\,a^3
=\,\left(\,E_{\rm{cell}}\,-\,T\,S_{\rm{cell}}\,\right)\,/\,a^3   ,
\end{equation}
where the free energy per cell $F_{\rm{cell}}$ can be written as
\begin{equation}
F_{\rm{cell}}=(E_{\rm{bulk}}+E_s+E_C)- T S_{\rm{cell}}=F_{\rm{bulk}}+E_s+E_C.
\end{equation}
The bulk energy $E_{\rm{bulk}}$, entropy $S_{\rm{cell}}$, and free energy
$F_{\rm{bulk}}$ are calculated by
\begin{eqnarray}
E_{\rm{bulk}} &=&\int_{\rm{cell}} \epsilon \left( \, n_n\left(r\right),
\, n_p\left(r\right), \, n_{\alpha}\left(r\right) \, \right) d^3r, \\
S_{\rm{cell}} &=&\int_{\rm{cell}} s \left(\, n_n\left(r\right),
\, n_p\left(r\right), \, n_{\alpha}\left(r\right) \, \right) d^3r, \\
F_{\rm{bulk}} &=&\int_{\rm{cell}} f \left( \, n_n\left(r\right),
\, n_p\left(r\right), \, n_{\alpha}\left(r\right) \, \right) d^3r.
\end{eqnarray}
Here $\epsilon \left( \, n_n\left(r\right),
\, n_p\left(r\right), \, n_{\alpha}\left(r\right) \, \right)$,
$s \left( \, n_n\left(r\right),
\, n_p\left(r\right), \, n_{\alpha}\left(r\right) \, \right)$,
and $f \left( \, n_n\left(r\right),
\, n_p\left(r\right), \, n_{\alpha}\left(r\right) \, \right)$
are the local energy density, entropy density, and free energy density
at the radius $r$, where the system can be considered as a mixed uniform matter
of neutrons, protons, and alpha-particles. These local densities are the sum
of the contributions from nucleons and alpha particles.
We use the RMF theory described in Section~\ref{sec:2.1} to calculate the nucleon
contribution, while the alpha-particles are treated as an ideal Boltzmann gas.
In general, the number density of alpha-particles is quite small, and therefore
the ideal-gas approximation is considered to be a reasonable approximation
for alpha-particles. We note that other treatments for alpha-particles
have been developed and used in nuclear astrophysics~\citep{sumi08,horo06}.
The free energy density of alpha-particles in the ideal-gas approximation
is given by
\begin{equation}
f_{\alpha} (n_{\alpha}) = -T\,n_{\alpha}\left[\ln (8n_Q/n_{\alpha})+1\right]
                          +n_{\alpha}\left(4M-B_{\alpha}\right),
\end{equation}
where $n_{\alpha}$ is the number density of alpha-particles, and we have used
the abbreviation $n_Q=\left[MT/(2\pi)\right]^{3/2}$.
The alpha-particle binding energy $B_{\alpha}=28.3$ MeV is
taken from~\citet{latt91}.
We have to take into account the volume of alpha-particle, otherwise the
alpha-particle fraction would become a large number at high density,
where the alpha-particle should actually disappear.
When we take into account the volume excluded by alpha-particles,
the free energy densities of nucleons and alpha-particles are given by
\begin{eqnarray}
f_N (n_n, n_p) &=& (1-u) f_N (\tilde{n}_n, \tilde{n}_p), \\
f_{\alpha} (n_{\alpha}) &=& (1-u) f_{\alpha} (\tilde{n}_{\alpha}),
\end{eqnarray}
where $u=n_{\alpha}v_{\alpha}$ is the fraction of space occupied by
alpha-particles with the effective volume of alpha-particle
$v_{\alpha}=24\,\rm{fm^{-3}}$ taken from~\citet{latt91}.
We denote the effective number density of neutrons, protons,
or alpha-particles as $\tilde{n}_i=n_i/(1-u)$ ($i=n$, $p$, or $\alpha$).
The inclusion of the volume excluded by alpha-particles
has negligible effect in the low-density region,
whereas it is necessary for the calculation at high density.

As for the surface energy term $E_s$ due to the inhomogeneity of the nucleon
distribution, we take the simple form
\begin{equation}
E_s=\int_{\rm{cell}} F_0 \mid \nabla \left( \, n_n\left(r\right)+
    n_p\left(r\right) \, \right) \mid^2 d^3r.
\end{equation}
The parameter $F_0=70 \, \rm{MeV\,fm^5}$ is determined by performing the Thomas--Fermi
calculation for finite nuclei so as to reproduce the gross properties of
nuclear masses, charge radii, and the beta stability line as described in
the Appendix of~\citet{oyam93}.

The Coulomb energy per cell $E_C$ is calculated using the Wigner--Seitz
approximation with an added correction term for the BCC lattice~\citep{oyam93}
\begin{equation}
\label{eq:ec}
E_C=\frac{1}{2}\int_{\rm{cell}} e \left[n_p\left(r\right)
+2n_{\alpha}\left(r\right)-n_e\right]\,\phi(r) d^3r
\,+\,\Delta E_C,
\end{equation}
where $\phi(r)$ represents the electrostatic potential calculated
in the Wigner--Seitz approximation,
$n_e$ is the electron number density of a uniform electron gas ($n_e=Y_p\,n_B$),
and $\Delta E_C$ is the correction term for the BCC lattice,
which is approximated as
\begin{equation}
\label{eq:dec}
\Delta E_C=C_{\rm{BCC}}\frac{(Z_{\rm{non}}e)^2} {a}.
\end{equation}
Here $a$ is the lattice constant,
$C_{\rm{BCC}}=0.006562$ is taken from~\citet{oyam93},
and $Z_{\rm{non}}$ is the non-uniform part of the charge number per cell
given by
\begin{equation}
Z_{\rm{non}}=\int_{0}^{R_p}
(n_p^{\rm{in}}-n_p^{\rm{out}}-2 n_{\alpha}^{\rm{out}})
\left[1-\left(\frac{r}{R_p}\right)^{t_p}\right]^3
4\pi r^2 dr.
\end{equation}
Because of the long-range nature of the Coulomb interaction,
$E_C$ is dependent on the lattice type.
This dependence has been extensively discussed in~\citet{oyam93}.
The system prefers the BCC lattice because it gives the lowest Coulomb energy.
In general, the Coulomb energy is dependent on the temperature
as discussed in~\citet{brav99} and~\citet{pote10}.
The dependence of $T$ in Equation~(\ref{eq:ec}) is contained in the particle
distributions. We assume that Equation~(\ref{eq:dec}), which was derived
at zero temperature in~\citet{oyam93}, remains valid at finite temperature.

Since we assume the lattice of nuclei with the Wigner--Seitz approximation,
we do not include a contribution from the translational energy of heavy nuclei.
This contribution is small in general and does not affect the general behavior
of the EOS. It may be noticeable only within the limited region of temperature,
where nuclei behave as gas without being dissociated into nucleons.
We note that this is different from the treatment by~\citet{latt91},
who included this term as a minor contribution.

%%%%%%%%%%%%%%%%%%%%%%%%%%%%%%%%%%%%%%%%%%%%%%%%%%%%%%%%%%%%%%%%%%%%%%%%%%%%%%%%
\section{Results}
\label{sec:3}

In this work, we construct the EOS tables covering a wide range of
temperature $T$, proton fraction $Y_p$, and baryon mass density $\rho_B$
for the use of core-collapse supernova simulations.
We present two sets of EOS tables, namely EOS2 and EOS3.
EOS2 takes into account only the nucleon degree of freedom,
while EOS3 includes additional contributions from $\Lambda$ hyperons.
In comparison with the earlier version (EOS1) described in~\citet{shen98b},
several improvements are made in EOS2 and EOS3 according to the requirements
of the users. We largely increase the number of $T$ points, and adopt a linear
$Y_p$ grid in EOS2 and EOS3 instead of the logarithmic $Y_p$ grid used in EOS1.
On the other hand, the numerical methods are improved to allow the
calculation with equal grid spacing for $\rho_B$, while it is only
approximately equal at high density in EOS1.
For a detailed comparison between the EOS tables discussed in this paper,
one can see Table 1 in Section~\ref{sec:1}.
In principle, the matter at extremely high temperatures and densities is
beyond the applicability of the RMF theory, but we still include
the results in these exotic regions since they are sometimes necessary
in astrophysical simulations.

%%%%%%%%%%%%%%%%%%%%%%%%%%%%%%%%%%%%%%%%
\subsection{Properties of matter without hyperons }
\label{sec:3.1}

In EOS2, we present results of matter without the inclusion of $\Lambda$ hyperons.
For each $T$, $Y_p$, and $\rho_B$, we determine the thermodynamically
favorable state that has the lowest free energy in the present model.
We perform the free energy minimization for both non-uniform matter
and uniform matter. Here the phase of heavy nuclei formed together with
free nucleons and alpha-particles is referred to as non-uniform matter,
while the phase of nucleons mixed with alpha-particles without
heavy nuclei is referred to as uniform matter.
For non-uniform matter, the minimization procedure is realized by using
the Thomas--Fermi method which includes eight independent parameters
as described in Section~\ref{sec:2.2}.
For uniform matter, we perform the minimization with respect to
converting two protons and two neutrons into an alpha-particle,
in which there is only one independent parameter.
By comparing the free energies of non-uniform matter and uniform matter,
we determine the most favorable state and estimate the phase transition between
non-uniform matter and uniform matter.

\begin{figure}[htb]
\includegraphics[bb=25 95 525 800, width=8.6 cm,clip]{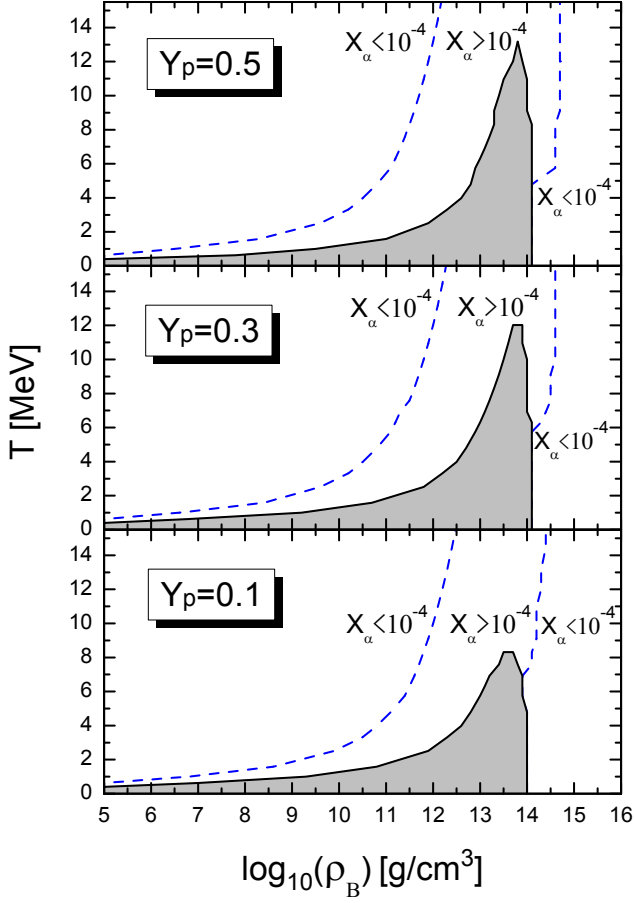}
\caption{Phase diagram of nuclear matter at $Y_p=0.1$, $0.3$,
and $0.5$ (bottom to top) in the $\rho_B$--$T$ plane.
The shaded region corresponds to the non-uniform matter phase where heavy
nuclei are formed. The dashed line is the boundary where the alpha-particle
fraction $X_{\alpha}$ changes between $X_{\alpha}<10^{-4}$ and $X_{\alpha}>10^{-4}$.
(A color version of this figure is available in the online journal.)}
\label{fig:TRho}
\end{figure}

\begin{figure}[htb]
\includegraphics[bb=25 95 525 800, width=8.6 cm,clip]{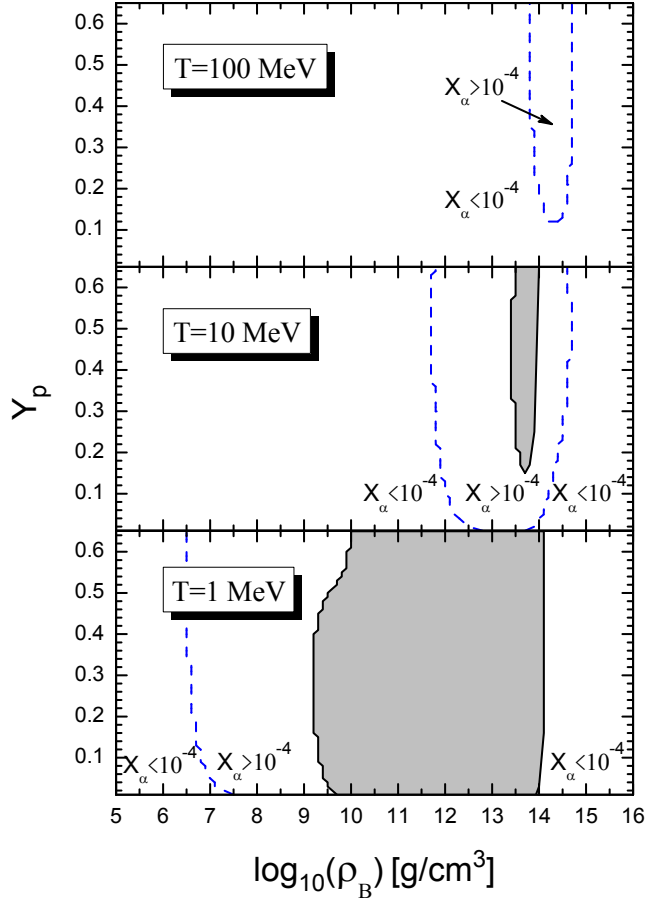}
\caption{Phase diagram of nuclear matter at $T=1$, $10$,
and $100$ MeV (bottom to top) in the $\rho_B$--$Y_p$ plane.
The shaded region corresponds to the non-uniform matter phase where heavy
nuclei are formed. The dashed line is the boundary where the alpha-particle
fraction $X_{\alpha}$ changes between $X_{\alpha}<10^{-4}$ and $X_{\alpha}>10^{-4}$.
The non-uniform matter phase does not exist at $T=100$ MeV (top).
(A color version of this figure is available in the online journal.)}
\label{fig:YpRho}
\end{figure}

We first discuss the phase diagram of nuclear matter at finite temperature.
It is known that the density of the phase transition between uniform matter
and non-uniform matter depends on both $T$ and $Y_p$. The non-uniform matter
phase can exist only in the low-temperature region ($T<14$ MeV).
In Figure~\ref{fig:TRho}, we show the phase diagram in the $\rho_B$--$T$
plane for $Y_p=0.1$, $0.3$, and $0.5$.
The shaded region corresponds to the non-uniform matter phase where heavy
nuclei are formed. The dashed line is the boundary where the alpha-particle
fraction $X_{\alpha}$ changes between $X_{\alpha}<10^{-4}$ and $X_{\alpha}>10^{-4}$.
It is shown that heavy nuclei can exist in the medium-density and low-temperature region.
The phase of nuclear matter at low density is a homogeneous nucleon gas with a small
fraction of alpha-particles, the heavy nuclei are formed at some medium densities
where the system can lower the free energy by forming heavy nuclei,
and it becomes uniform matter as the density increases
beyond $\sim 10^{14.2}\,\rm{g\,cm^{-3}}$.
It is seen that the starting density of the non-uniform matter phase depends
on $T$ strongly, while the ending density is nearly independent of $T$.
As the temperature increases, the density range of the non-uniform matter phase
becomes narrower, and it disappears completely for $T > 14$ MeV.
In Figure~\ref{fig:YpRho}, we show the phase diagram in the $\rho_B$--$Y_p$
plane for $T=1$, $10$, and $100$ MeV.
The shaded region corresponds to the non-uniform matter phase,
while the dashed line is the boundary between $X_{\alpha}<10^{-4}$
and $X_{\alpha}>10^{-4}$. It is shown that the dependence of the boundary on $Y_p$
is relatively weak except at very small values of $Y_p$.
In the case of $T=10$ MeV (middle panel),
the non-uniform matter phase disappears at lower $Y_p$ because it is difficult
to form heavy nuclei with smaller values of $Y_p$.
The alpha-particles still exist at $T=100$ MeV, and $X_{\alpha}$ reaches to be
more than $10^{-4}$ at some medium densities for $Y_p>0.1$ as shown in the top
panel of Figure~\ref{fig:YpRho}.
We note that the phase diagram of nuclear matter at zero temperature
has been discussed in~\citet{shen98a,shen98b}.

\begin{figure}[htb]
\includegraphics[bb=20 95 525 800, width=8.6 cm,clip]{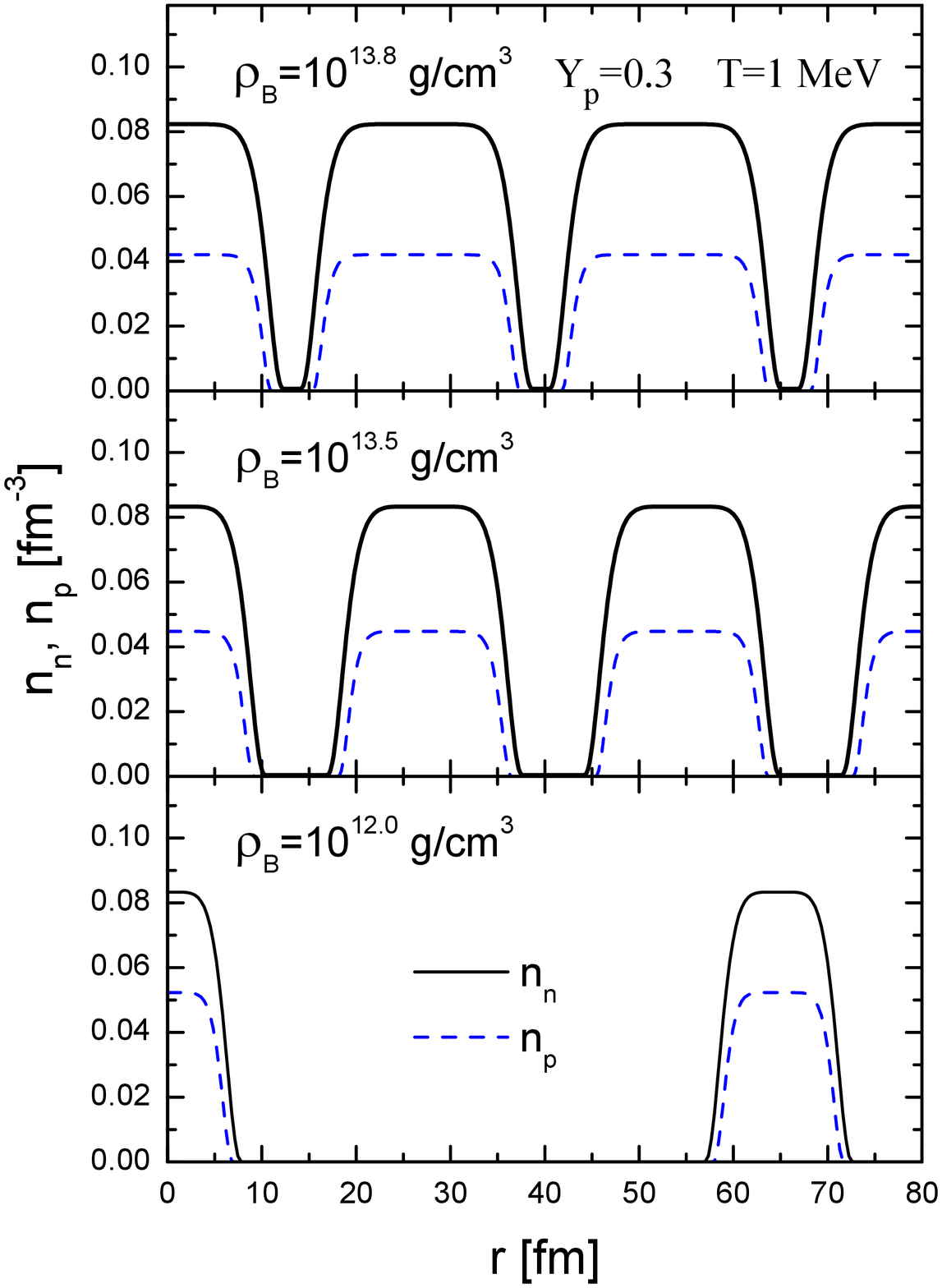}
\caption{Density distributions of neutrons (solid) and protons (dashed)
along the straight line joining the centers of the nearest nuclei in
the BCC lattice for the case of $T=1$ MeV and $Y_p=0.3$ at
$\rho_B=10^{12}$, $10^{13.5}$, and $10^{13.8}\,\rm{g\,cm^{-3}}$ (bottom to top).
(A color version of this figure is available in the online journal.)}
\label{fig:NM_T1}
\end{figure}

\begin{figure}[htb]
\includegraphics[bb=20 95 525 600, width=8.6 cm,clip]{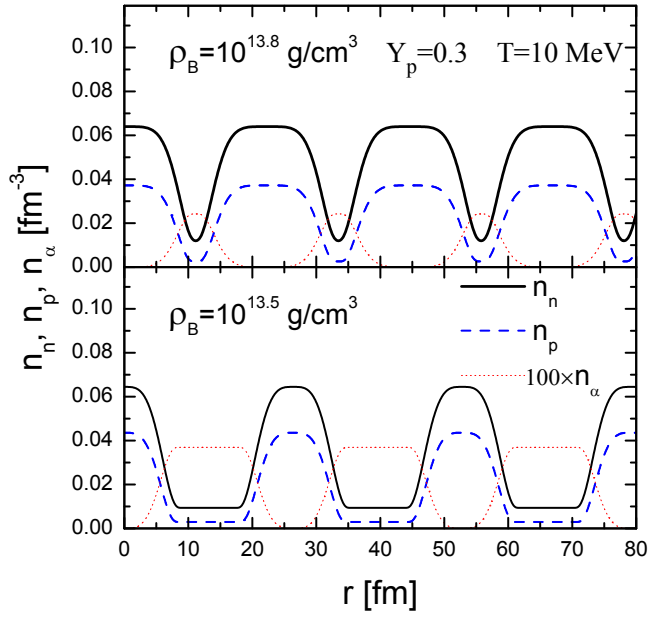}
\caption{Density distributions of neutrons (solid), protons (dashed),
and alpha-particles (dotted) along the straight line joining the
centers of the nearest nuclei in the BCC lattice for the case
of $T=10$ MeV and $Y_p=0.3$ at $\rho_B=10^{13.5}$
and $10^{13.8}\,\rm{g\,cm^{-3}}$ (bottom to top).
(A color version of this figure is available in the online journal.)}
\label{fig:NM_T10}
\end{figure}

\begin{figure}[htb]
\includegraphics[bb=25 95 525 600, width=8.6 cm,clip]{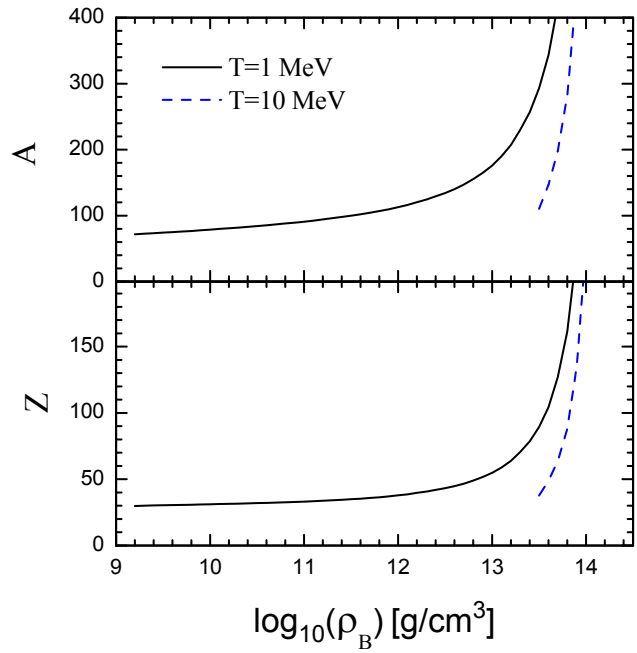}
\caption{Nuclear mass number $A$ (top) and charge number $Z$ (bottom)
as a function of the baryon mass density $\rho_B$
for $Y_p=0.3$ at $T=1$ MeV (solid) and $T=10$ MeV (dashed).
(A color version of this figure is available in the online journal.)}
\label{fig:AZ}
\end{figure}

In the non-uniform matter, nucleon distributions are determined by
minimizing the free energy density with respect to the independent parameters
in the Thomas--Fermi approximation. The heavy nuclei are assumed to form a BCC
lattice in order to minimize the Coulomb lattice energy.
In Figure~\ref{fig:NM_T1}, we show the nucleon distributions along
the straight line joining the centers of the nearest nuclei in
the BCC lattice for the case of $T=1$ MeV and $Y_p=0.3$.
It is found that the nuclei become heavier and get closer with each other
as the density increases. In principle, there exist free nucleons
and alpha-particles outside heavy nuclei at finite temperature, but
their densities are too small to be observed in Figure~\ref{fig:NM_T1}.
We show in Figure~\ref{fig:NM_T10} the same quantities as in Figure~\ref{fig:NM_T1}
but for the case of $T=10$ MeV and $Y_p=0.3$.
We note that the starting density of the non-uniform matter phase in this case is
$10^{13.5}\,\rm{g\,cm^{-3}}$, so the heavy nucleus does not exist at $\rho_B=10^{12}\,\rm{g\,cm^{-3}}$
for $T=10$ MeV which is different from the case of $T=1$ MeV
as shown in the bottom panel of Figure~\ref{fig:NM_T1}.
Comparing with the top and middle panels of Figure~\ref{fig:NM_T1},
it is seen that there are much more free nucleons and alpha-particles outside
nuclei, while the nucleons inside nuclei get less.
This is because the role of entropy becomes more important at higher temperature
and the free energy can be lowered if the nucleons are freed from nuclei.
In Figure~\ref{fig:AZ}, we plot the nuclear mass number $A$
and charge number $Z$ as a function of the baryon mass density $\rho_B$
for $Y_p=0.3$ at $T=1$ MeV and $T=10$ MeV.
It is shown that $A$ and $Z$ have relatively weak dependence on $\rho_B$ at lower
density and they increase rapidly just before the phase transition
at $\rho_B \sim 10^{14}\,\rm{g\,cm^{-3}}$.
For the same $\rho_B$ and $Y_p$, the values of $A$ and $Z$ at $T=10$ MeV are
much smaller than those at $T=1$ MeV. This is because more nucleons
are freed from nuclei at higher temperature, and it is eventually impossible
to form heavy nuclei for $T > 14$ MeV in the present model.

\begin{figure}[htb]
\includegraphics[bb=25 95 525 800, width=8.6 cm,clip]{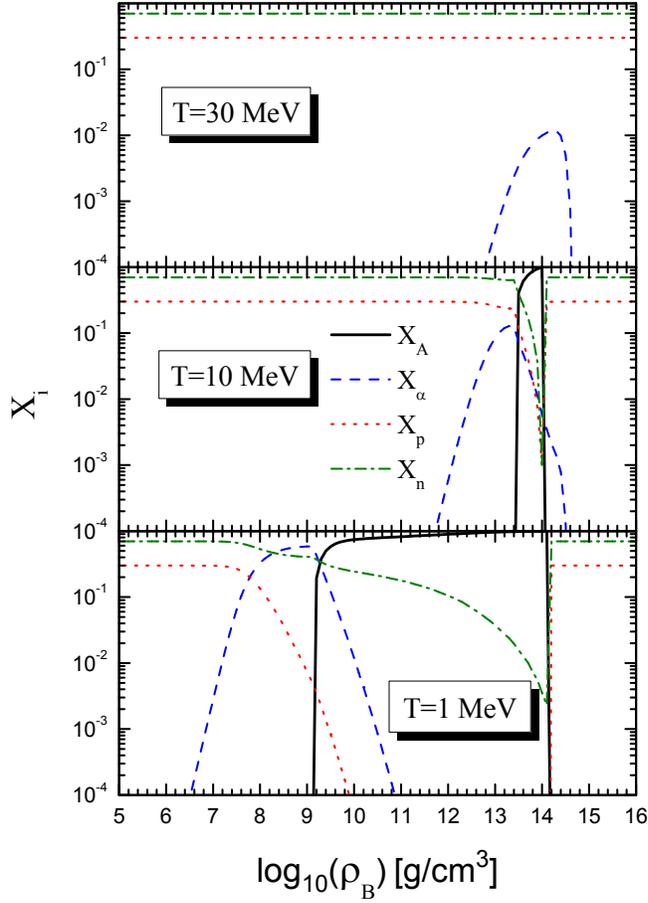}
\caption{Fraction of neutrons (dash-dotted), protons (dotted), alpha-particles (dashed),
and heavy nuclei (solid) as a function of the baryon mass density $\rho_B$
for $Y_p=0.3$ at $T=1$, $10$, and $30$ MeV (bottom to top).
(A color version of this figure is available in the online journal.)}
\label{fig:XiRho2}
\end{figure}

In Figure~\ref{fig:XiRho2}, we show the fraction of neutrons, protons, alpha-particles,
and heavy nuclei as a function of the baryon mass density $\rho_B$ for $Y_p=0.3$
at $T=1$, $10$, and $30$ MeV. At low density, the matter is a uniform gas of
neutrons and protons with a small fraction of alpha-particles.
The alpha-particle fraction $X_{\alpha}$ increases with increasing $\rho_B$,
but the formation of heavy nuclei at low temperatures causes a rapid drop of
$X_{\alpha}$, $X_p$, and $X_n$, which is due to the fact that heavy nuclei
use up most of the nucleons in non-uniform matter.
When the density increases beyond $\sim 10^{14.2}\,\rm{g\,cm^{-3}}$,
the heavy nuclei dissolve and the matter becomes uniform.
The alpha-particles may exist up to $\rho_B \sim 10^{14.6}\,\rm{g\,cm^{-3}}$
where the volume excluded by alpha-particles plays an important role
that it is unfavorable to have alpha-particles in the uniform matter
at such high density. For $T > 14$ MeV, the heavy nuclei cannot be formed,
but there are finite values of $X_{\alpha}$, especially at medium densities
as shown in the top panel of Figure~\ref{fig:XiRho2}.
The alpha-particle fraction decreases with increasing temperature, and we find
that $X_{\alpha}$ at $T=100$ MeV is of the order of $10^{-4}$ at some medium densities.
The alpha-particle fraction may be significantly affected if the alpha-particle
binding energy $B_{\alpha}$ is dependent on the density and temperature
as discussed in~\citet{ropk05}. For simplicity, we neglect this dependence
and take $B_{\alpha}=28.3$ MeV in the present model.
Since we treat the uniform matter and non-uniform matter consistently using
the same RMF theory, all the resulting thermodynamic quantities are consistent
and smooth in the whole range. We will discuss the thermodynamic quantities
in EOS2 and compare with those in EOS3 in the next section.

%%%%%%%%%%%%%%%%%%%%%%%%%%%%%%%%%%%%%%%%
\subsection{Properties of matter with the inclusion of $\Lambda$ hyperons}
\label{sec:3.2}

In this section, we discuss results of matter with the inclusion of $\Lambda$
hyperons given in EOS3. It is found that the contribution from $\Lambda$
hyperons is negligible in the low-temperature and low-density region.
In fact, at zero temperature $\Lambda$ hyperons appear only at high density
when the equilibrium condition $\mu_n=\mu_{\Lambda}$ could be satisfied.
At low temperature, such as $T=10$ MeV, the $\Lambda$ fraction $X_{\Lambda}$
is smaller than $10^{-5}$ at densities below normal nuclear matter density.
For simplicity, we take into account the contribution from $\Lambda$ hyperons
only when $X_{\Lambda}$ is larger than $10^{-5}$.
In the non-uniform matter phase, $X_{\Lambda}$ is quite small,
therefore we neglect the contribution from $\Lambda$ hyperons
in the Thomas--Fermi calculation.

\begin{figure}[htb]
\includegraphics[bb=25 95 525 800, width=8.6 cm,clip]{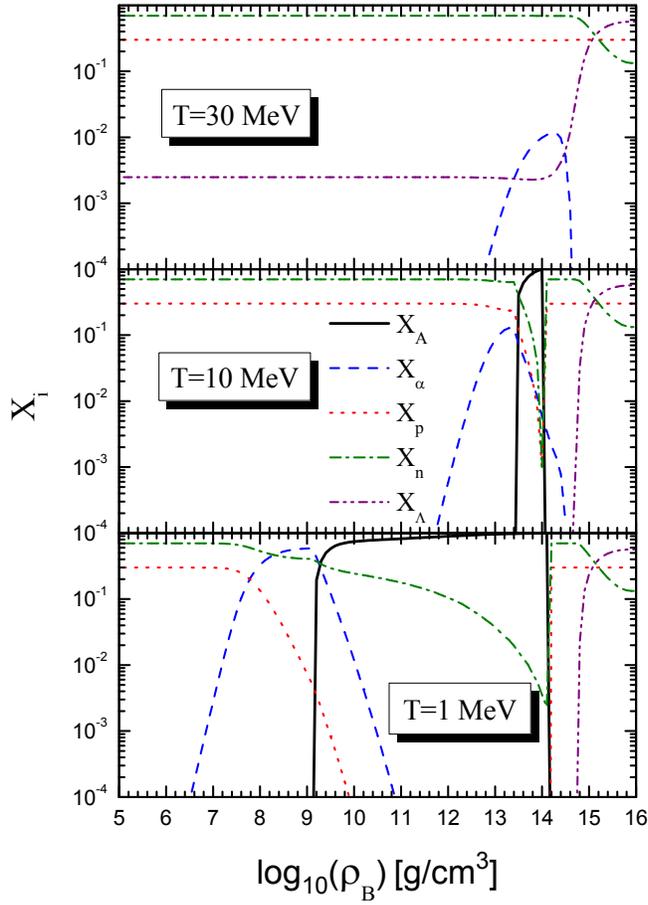}
\caption{Same as Figure~\ref{fig:XiRho2}, but with the inclusion of $\Lambda$ hyperons.
(A color version of this figure is available in the online journal.)}
\label{fig:XiRho3}
\end{figure}

In Figure~\ref{fig:XiRho3}, we show the fraction $X_i$ ($i=\Lambda$, $n$, $p$,
$\alpha$, or heavy nuclei $A$) as a function of $\rho_B$ for $Y_p=0.3$
at $T=1$, $10$, and $30$ MeV. For lower temperatures, such as $T=1$ MeV (bottom panel)
and $T=10$ MeV (middle panel), there is a significant fraction of $\Lambda$ hyperons
only at high density. It is shown that $X_{\Lambda}$ increases with increasing $\rho_B$,
which causes a decrease of $X_n$. We note that $X_p$, which is equal to $Y_p$ in the uniform
matter at high density, has been fixed to be $0.3$ in this figure.
For $T=30$ MeV (top panel), $X_{\Lambda}$ is of the order of $10^{-3}$ at low density,
and increases rapidly at high density.
Comparing with the top panel of Figure~\ref{fig:XiRho2},
it is seen that the inclusion of $\Lambda$ hyperons does not affect $X_{\alpha}$
and $X_p$ significantly. We find the contribution from $\Lambda$ hyperons increases
with increasing temperature. At $T=100$ MeV, $X_{\Lambda}$ is about 12\% at low density
for the case of $Y_p=0.3$.  On the other hand, $X_{\Lambda}$ decreases with increasing $Y_p$
at fixed $\rho_B$ and $T$, since there are less neutrons at higher $Y_p$
to realize the equilibrium condition $\mu_n=\mu_{\Lambda}$.

\begin{figure}[htb]
\includegraphics[bb=15 95 525 800, width=8.6 cm,clip]{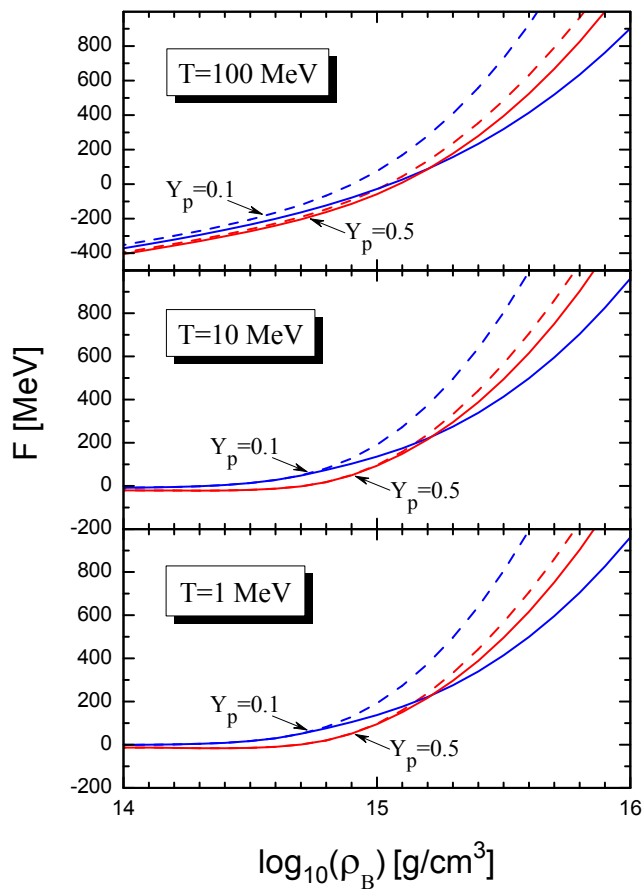}
\caption{Free energy per baryon $F$ as a function of the baryon mass
density $\rho_B$ with $Y_p=0.1$ (blue) and $0.5$ (red)
at $T=1$, $10$, and $100$ MeV (bottom to top).
The results with $\Lambda$ hyperons given in EOS3 are shown by solid lines,
while those without $\Lambda$ hyperons given in EOS2 are displayed by dashed lines.
(A color version of this figure is available in the online journal.)}
\label{fig:F}
\end{figure}

\begin{figure}[htb]
\includegraphics[bb=15 95 525 800, width=8.6 cm,clip]{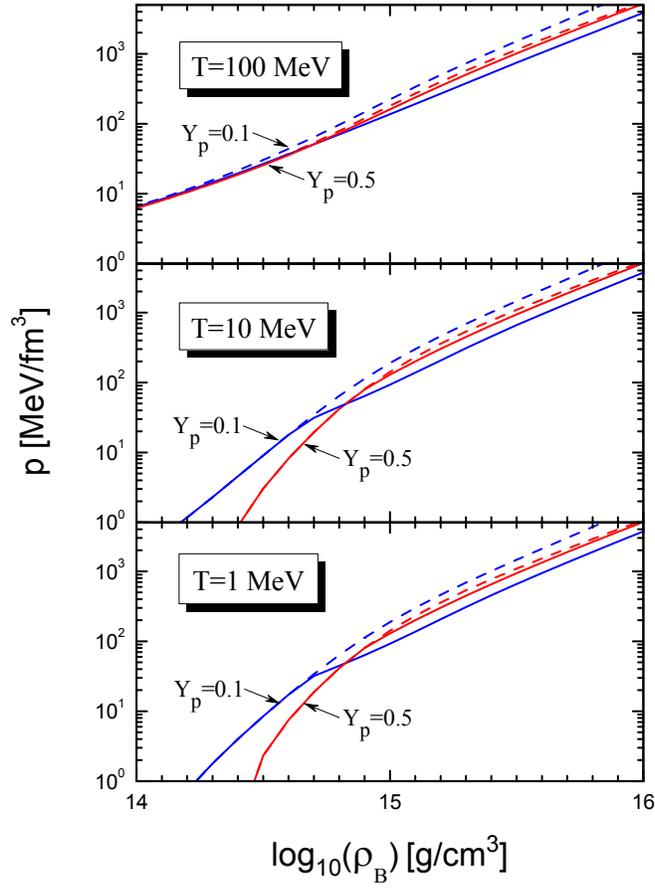}
\caption{Same as Figure~\ref{fig:F}, but for the pressure $p$.
(A color version of this figure is available in the online journal.)}
\label{fig:p}
\end{figure}

\begin{figure}[htb]
\includegraphics[bb=15 95 525 800, width=8.6 cm,clip]{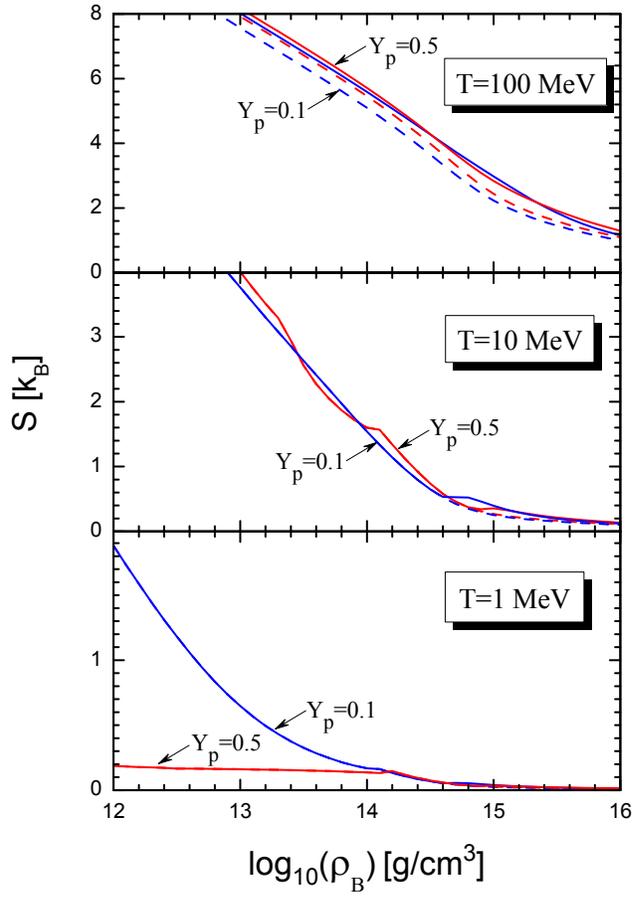}
\caption{Same as Figure~\ref{fig:F}, but for the entropy per baryon $S$.
(A color version of this figure is available in the online journal.)}
\label{fig:S}
\end{figure}

We now discuss the thermodynamic quantities in EOS3 and compare with those in EOS2.
Here we mainly discuss their properties at high density where there are noticeable
differences between the results with and without $\Lambda$ hyperons. As for the
behavior of thermodynamic quantities at low density, one can refer to our earlier
work~\citep{shen98a,shen98b}.
In Figure~\ref{fig:F}, we show the free energy per baryon $F$ as a function of the
baryon mass density $\rho_B$ with $Y_p=0.1$ and $0.5$ at $T=1$, $10$, and $100$ MeV.
The results with $\Lambda$ hyperons given in EOS3 are shown by solid lines,
while those without $\Lambda$ hyperons given in EOS2 are displayed by dashed lines.
It is found that the inclusion of $\Lambda$ hyperons can lower the free energy
and this effect increases with increasing $\rho_B$. However, this effect decreases
with increasing $Y_p$ as shown in Figure~\ref{fig:F}, which is due to the fact that
$X_{\Lambda}$ decreases with increasing $Y_p$.
We show in Figure~\ref{fig:p} the pressure $p$ as a function of $\rho_B$
with $Y_p=0.1$ and $0.5$ at $T=1$, $10$, and $100$ MeV.
The pressure is calculated from the derivative of the free energy as described
in Appendix A. Therefore, the effect of $\Lambda$ hyperons on the pressure is
similar to the one observed in Figure~\ref{fig:F}. It is obvious that the
inclusion of $\Lambda$ hyperons tends to soften the EOS at high density.
In Figure~\ref{fig:S}, we show the entropy per baryon $S$ as a function of
$\rho_B$ with $Y_p=0.1$ and $0.5$ at $T=1$, $10$, and $100$ MeV.
At $T=1$ MeV (bottom panel) it is hard to see the difference between the
results with and without $\Lambda$ hyperons, while at $T=10$ MeV (middle panel)
there are small differences at high density.
For the case of $T=100$ MeV (top panel), the effect of $\Lambda$ hyperons
can be seen in the whole range of density,
since $X_{\Lambda}$ reaches $\sim 16\%$ ($\sim 9\%$)
at low density for $Y_p=0.1$ ($0.5$) at $T=100$ MeV.
In general, the inclusion of $\Lambda$ hyperons tends to increase the entropy.
In~\citet{shen98a,shen98b}, we have discussed the effect of the formation of
nuclei on the entropy, which has a strong $Y_p$ dependence.
The behavior of $S$ at $\rho_B < 10^{14}\,\rm{g\,cm^{-3}}$ in the middle and bottom
panels of Figure~\ref{fig:S} is due to the formation of heavy nuclei in non-uniform matter.

For neutron star matter at zero temperature, the inclusion of $\Lambda$ hyperons
leads to a softening of the EOS of neutron star matter
and thereby a corresponding reduction in the maximum mass of neutron stars.
We use EOS3 to calculate the neutron star properties and find that
the maximum mass of neutron stars is about 1.75 $M_{\odot}$,
whereas the value for EOS1 and EOS2 is 2.18 $M_{\odot}$.
According to the recent measurement of PSR J1614-2230
($1.97\, M_{\odot}$;~\citet{demo10}), EOS3 seems to be too soft
due to the inclusion of $\Lambda$ hyperons,
while EOS1 and EOS2 are compatible with the observation
of PSR J1614-2230.
This is a common difficulty for EOSs with the inclusion of hyperons.
So far, there are large uncertainties in the properties of matter at high density,
which are crucial for determining the maximum mass of neutron stars.

%%%%%%%%%%%%%%%%%%%%%%%%%%%%%%%%%%%%%%%%%%%%%%%%%%%%%%%%%%%%%%%%%%%%%%%%%%%%%%%%
\section{Summary}
\label{sec:4}

In this paper, we have presented two sets of the EOS tables (EOS2 and EOS3)
covering a wide range of temperature $T$, proton fraction $Y_p$,
and baryon mass density $\rho_B$ for the use of core-collapse
supernova simulations. The difference between EOS2 and EOS3
is that only the nucleon degree of freedom is taken into account
in EOS2, while EOS3 includes additional contributions from $\Lambda$ hyperons.
In comparison with the earlier version (EOS1) described in~\citet{shen98b},
several improvements have been made in EOS2 and EOS3 according to the
requirements of the users. We have largely increased the number of $T$ points,
and adopted a linear $Y_p$ grid in EOS2 and EOS3 instead of the logarithmic $Y_p$
grid used in EOS1. In addition, we have performed the calculation with
equal grid spacing in the whole range of density, while it is only
approximately equal at high density in EOS1.
We have presented a detailed comparison between these EOSs
in Table 1.

We have employed the RMF theory with nonlinear $\sigma$ and $\omega$
terms, which can provide an excellent description of nuclear matter
and finite nuclei including unstable ones.
The Thomas--Fermi approximation in combination with assumed nucleon
distribution functions and a free energy minimization has been adopted
to describe the non-uniform matter that is modeled as a mixture of a single species
of heavy nuclei, alpha particles, and free nucleons outside of nuclei.
The RMF results have been taken as input in the Thomas--Fermi calculation,
so the treatments of non-uniform matter and uniform matter
in this EOS are sufficient to obtain the table in a consistent manner.
It would be preferable to treat the mixture of nuclei as recently done
in~\citet{hemp10},~\citet{furu11}, and~\citet{blin11} for the detailed treatment of
electron captures on nuclei in supernova core.
However, it is beyond the scope of the current update of the Shen EOS table,
which has been routinely used in astrophysical simulations.

We have included the contribution from $\Lambda$ hyperons in EOS3,
when the $\Lambda$ fraction is larger than $10^{-5}$.
It is based on the consideration that the $\Lambda$ hyperon
is most likely to occur in dense matter among all possible
non-nucleonic degrees of freedom because it is the lightest
hyperon with an attractive potential in nuclear matter.
The potential depth of $\Lambda$ in nuclear matter is estimated
to be around $-30$ MeV from the experimental binding energies of
single-$\Lambda$ hypernuclei.
Several recent observations of double-$\Lambda$ hypernuclei
suggest that the effective $\Lambda\Lambda$ interaction should be
weakly attractive.
The RMF theory with nonlinear $\sigma$ and $\omega$ terms has been
extended to include hyperons, and it can provide a reasonable
description of single- and double-$\Lambda$ hypernuclei~\citep{shen06}.
We have not included contributions from other hyperons, such as $\Sigma$
and $\Xi$, due to their relatively high threshold densities and
lack of available experimental data.
In EOS3, we have performed a consistent calculation including
$\Lambda$ hyperons for the entire table, whereas the EOS tables of
high density matter were connected with EOS1 at low density
in the simple procedure described in Section 2.3 of~\citet{ishi08},
so EOS3 is considered to be constructed in a consistent manner
for the inclusion of $\Lambda$ hyperons.
We have examined the effect of $\Lambda$ hyperons on the properties
of dense matter. The contribution from $\Lambda$ hyperons is negligible
in the low-temperature and low-density region, whereas it tends to soften
the EOS at high density.

In principle, the matter at extremely high temperatures and densities
is beyond the applicability of the RMF theory, but we still include
the results in these exotic regions since they are sometimes necessary
in astrophysical simulations. It will continue to be a challenge
for nuclear physics to provide realistic EOS, which should be supported
by microscopic theory and available experimental data,
for use in astrophysical studies.

%%%%%%%%%%%%%%%%%%%%%%%%%%%%%%%%%%%%%%%%%%%%%%%%%%%%%%%%%%%%%%%%%%%%%%%%%%%%%%%%
\acknowledgments

We thank C. D. Ott and T. Fischer for helpful suggestions and fruitful discussions.
K.S. expresses his thanks to K. Nakazato, A. Ohnishi, H. Suzuki, and S. Yamada
for practical discussions on the development of the EOS tables for simulations.
H.S. and K.S. thank the organizers of the MICRA2011 workshop
hosted by the Perimeter Institute.
This work was partially supported by the National Natural Science
Foundation of China (grant nos. 10675064 and 11075082),
the Grant-in-Aid for Scientific Research on Innovative Areas (nos. 20105004 and 20105005),
and the Grant-in-Aid for the Scientific Research (no. 22540296)
from the Ministry of Education, Culture, Sports, Science and Technology (MEXT) of Japan.
This work for numerical simulations of core-collapse supernovae
is partially supported by the HPCI Strategic Program of MEXT in Japan,
which assists K.S. to use the supercomputers at Yukawa Institute for
Theoretical Physics (YITP) in Kyoto University,
Research Center for Nuclear Physics (RCNP) in Osaka University,
the University of Tokyo, Japan Atomic Energy Agency (JAEA) and
High Energy Accelerator Research Organization (KEK).

%%%%%%%%%%%%%%%%%%%%%%%%%%%%%%%%%%%%%%%%%%%%%%%%%%%%%%%%%%%%%%%%%%%%%%%%%%%%%%%%
\appendix

\section{Physical quantities in the EOS tables}

The EOS tables are available on the Web at \\
{\it http://physics.nankai.edu.cn/grzy/shenhong/EOS/index.html}, \\
{\it http://www.rcnp.osaka-u.ac.jp/$^{\sim}$shen/}, \\
{\it http://user.numazu-ct.ac.jp/$^{\sim}$sumi/eos/index.html}. \\
The table under the name \textquotedblleft $\ast$.tab\textquotedblright\ is
the main EOS table, while \textquotedblleft $\ast$.t00\textquotedblright\
and \textquotedblleft $\ast$.yp0\textquotedblright\ are those for $T=0$ and
$Y_p=0$, respectively.
EOS2 and EOS3 are presented in electronic tables, Tables 3--8, in this paper.
We present the main EOS table of EOS2 (EOS3) in Table 3 (Table 6),
which is 149 MB (167 MB) in size.
Table 4 (Table 7) is the one of EOS2 (EOS3) at $T=0$  which is 1.6 MB (1.8 MB),
while Table 5 (Table 8) is the one for $Y_p=0$ which is 2.3 MB (2.6 MB).
The tables are written in the order of increasing $T$.
In the tables on the Web, the values of $\log_{10}(T)$ and $T$ (in MeV)
are given at the beginning of each block, and the blocks with different $T$ are
divided by the string of characters \textquotedblleft cccccccccccc\textquotedblright.
In Tables 3--8 of this paper, $\log_{10}(T)$ and $T$ are set in columns 1 and 2.
We note that the entry for $\log_{10}(T)$ is set to $-100$
in the tables for the case of $T=0$.

For each $T$, we present the results in the order of
increasing $Y_p$ and $\rho_B$.
The quantities in one line of Tables 3--8 are defined as follows.
\begin{itemize}
\item
1. Logarithm of temperature: $\log_{10}(T)$ [MeV].
\item
2. Temperature: $T$ [MeV].
\item
3. Logarithm of baryon mass density: $\log_{10}(\rho_B)$ [$\rm{g\,cm^{-3}}$].
\item
4. Baryon number density: $n_B$ [$\rm{fm^{-3}}$]. \\
The baryon number density is related to the baryon mass density as
$\rho_B=m_{u} n_B$ with $m_{u}=931.494$ MeV
being the atomic mass unit taken from~\citet{amsl08}.
\item
5. Proton fraction: $Y_p$.  \\
The proton fraction $Y_p$ of uniform matter is defined by
\begin{equation}
Y_p=\frac{n_p+2n_{\alpha}}{n_B}
   =\frac{n_p+2n_{\alpha}}{n_p+n_n+n_{\Lambda}+4n_{\alpha}},
\end{equation}
where $n_p$, $n_n$, $n_{\Lambda}$, and $n_{\alpha}$ are the number
density of protons, neutrons, $\Lambda$ hyperons, and alpha-particles, respectively.
$n_B$ is the total baryon number density.
For non-uniform matter case, $Y_p$ is the average proton fraction defined by
\begin{equation}
Y_p=\frac{N_p}{N_B},
\end{equation}
where $N_p$ and $N_B$ are the proton and baryon numbers per cell given by
\begin{eqnarray}
N_p &=& \int_{\rm{cell}} \left[\, n_p\left(r\right) + 2n_{\alpha}\left(r\right)
                      \,\right] d^3r, \\
N_B &=& \int_{\rm{cell}} \left[\, n_n\left(r\right) +
                         n_p\left(r\right) + 4n_{\alpha}\left(r\right)
                      \,\right] d^3r.
\end{eqnarray}
Here, $n_p(r)$ and $n_n(r)$ are the proton and neutron distributions
given by Equation~(\ref{eq:nitf}), and $n_{\alpha}(r)$ is the alpha-particle
distribution given by Equation~(\ref{eq:na}).
Because the $\Lambda$ fraction is very small in the non-uniform matter phase,
we neglect the contribution from $\Lambda$ hyperons in this case.
\item
6. Free energy per baryon: $F$  [MeV]. \\
The free energy per baryon is defined relative to
the free nucleon mass $M=938$ MeV in the TM1 parameter set as
\begin{equation}
F=\frac{f}{n_B}-M.
\end{equation}
\item
7. Internal energy per baryon: $E_{\rm{int}}$  [MeV]. \\
The internal energy per baryon is defined relative to
the atomic mass unit $m_{u}=931.494$ MeV as
\begin{equation}
E_{\rm{int}}=\frac{\epsilon}{n_B}-m_{u}.
\end{equation}
\item
8. Entropy per baryon: $S$  [$k_B$]. \\
The entropy per baryon is related to the entropy density via
\begin{equation}
S=\frac{s}{n_B}.
\end{equation}
\item
9. Mass number of the heavy nucleus: $A$.   \\
The mass number of the heavy nucleus is defined by
\begin{equation}
\label{eq:AN}
A=\int_{0}^{R_A}  \left[\, n_n\left(r\right) + n_p\left(r\right)
  \,\right] 4\pi r^2 dr,
\end{equation}
where $R_{A}$ is the maximum of $R_p$ and $R_n$, which is considered as
the boundary of the heavy nucleus.
\item
10. Charge number of the heavy nucleus: $Z$.   \\
The charge number of the heavy nucleus is defined by
\begin{equation}
Z=\int_{0}^{R_A}  \, n_p\left(r\right) \, 4\pi r^2 dr.
\end{equation}
\item
11. Effective nucleon mass: $M^{*}_N$ [MeV]. \\
The effective nucleon mass is obtained in the RMF theory for uniform matter.
In the non-uniform matter phase, the effective nucleon mass is a function of space
due to inhomogeneity of the nucleon distribution, so it is meaningless
to list the effective nucleon mass for non-uniform matter.
We replace the effective nucleon mass $M^{*}_N$ by the free nucleon mass $M$
in the non-uniform matter phase.
\item
12. Free neutron fraction: $X_n$.  \\
The free neutron fraction is given by
\begin{equation}
X_n=(n_n^{\rm{out}} V^{\rm{out}})/(n_B V_{\rm{cell}}),
\end{equation}
where $V_{\rm{cell}}=a^3=4 \pi R_{\rm{cell}}^3 / 3$ is the cell volume,
$V^{\rm{out}}=V_{\rm{cell}}-4 \pi R_{A}^3 / 3$ is the volume outside
the nucleus, $n_n^{\rm{out}}$ is the number density of free neutrons
outside the nucleus, and $n_B$ is the average baryon number density.
\item
13. Free proton fraction: $X_p$.  \\
The free proton fraction is given by
\begin{equation}
X_p=(n_p^{\rm{out}} V^{\rm{out}})/(n_B V_{\rm{cell}}),
\end{equation}
where $n_p^{\rm{out}}$ is the number density of free protons
outside the nucleus.
\item
14. Alpha-particle fraction: $X_{\alpha}$.  \\
The alpha-particle fraction is defined by
\begin{equation}
X_{\alpha}=4N_{\alpha}/(n_B V_{\rm{cell}}),
\end{equation}
where $N_{\alpha}$ is the alpha-particle number per cell obtained by
\begin{equation}
N_{\alpha}=\int_{\rm{cell}} n_{\alpha}\left(r\right) d^3r
\end{equation}
and $n_{\alpha}(r)$ is the alpha-particle distribution
given by Equation~(\ref{eq:na}).
\item
15. Heavy nucleus fraction: $X_A$.  \\
The heavy nucleus fraction is defined by
\begin{equation}
X_A=A/(n_B V_{\rm{cell}}),
\end{equation}
where $A$ is the mass number of the heavy nucleus
as defined in Equation~(\ref{eq:AN}).
\item
16. Pressure: $p$ [$\rm{MeV\,fm^{-3}}$]. \\
The pressure is calculated through the
thermodynamic relation
\begin{equation}
p  =\left[\,n_B^2 (\partial F/\partial n_B) \,\right]_{T,Y_p}.
\end{equation}
\item
17. Chemical potential of the neutron: $\mu_n$ [MeV]. \\
The chemical potential of the neutron relative to
the free nucleon mass $M$ is calculated through
the thermodynamic relation
\begin{equation}
\mu_n=\left[\,\partial (n_B F) /\partial n_{n+\Lambda}  \,\right]_{T,n_p},
\end{equation}
where $n_{n+\Lambda}=(1-Y_p)\,n_B$ is the sum of the neutron and $\Lambda$
densities.
\item
18. Chemical potential of the proton: $\mu_p$ [MeV]. \\
The chemical potential of the proton relative to
the free nucleon mass $M$ is calculated through
the thermodynamic relation
\begin{equation}
\mu_p=\left[\,\partial (n_B F) /\partial n_p  \,\right]_{T,n_{n+\Lambda}}.
\end{equation}
\item
19. Effective $\Lambda$ mass: $M^{*}_{\Lambda}$ [MeV]. \\
The effective $\Lambda$ mass is obtained in the RMF theory for uniform matter.
We replace the effective $\Lambda$ mass by the free $\Lambda$ mass $M_{\Lambda}=1115.7$ MeV
when the $\Lambda$ hyperon is not taken into account.
\item
20. $\Lambda$ fraction: $X_{\Lambda}$. \\
The $\Lambda$ fraction is given by
\begin{equation}
X_{\Lambda}=n_{\Lambda}/n_B,
\end{equation}
where $n_{\Lambda}$ is the number density of $\Lambda$ hyperons in the uniform matter phase.
We note that $X_{\Lambda}=0$ is adopted in the non-uniform matter phase.
\end{itemize}

%%%%%%%%%%%%%%%%%%%%%%%%%%%%%%%%%%%%%%%%
\section{Checks on the EOS tables}

We have done the following checks on the EOS tables.
\begin{itemize}
\item
1. Consistency of the fractions:
\begin{equation}
        X_p+X_n+X_{\Lambda}+X_{\alpha}+X_A=1.
\end{equation}
\item
2. Consistency of the relation between $F$, $E_{\rm{int}}$, and $S$:
\begin{equation}
        F=E_{\rm{int}}-TS +m_{u} -M.
\end{equation}
\item
3. Consistency of the thermodynamic quantities:
\begin{equation}
        F=\mu_n (1-Y_p)+\mu_p Y_p-\frac{p}{n_B}.
\end{equation}
\end{itemize}
In general, these consistency relations can be satisfied
within a few thousandths.
The physical constants used in this study are taken from~\citet{amsl08}.

%%%%%%%%%%%%%%%%%%%%%%%%%%%%%%%%%%%%%%%%%%%%%%%%%%%%%%%%%%%%%%%%%%%%%%%%%%%%%%%%

%%%%%%%%%%%%%%%%%%%%%%%%%%%%%%%%%%%%%%%%%%%%%%%%%%%%%%%%%%%%%%%%%%%%%%%%%%%%%%%%
\clearpage
%%%%%%%%%%%%%%
\begin{deluxetable}{ccccc}
\tabletypesize{\scriptsize}
\tablecaption{Comparison between the EOS tables discussed in this paper
\label{tab:1}}
\tablewidth{0cm}
\tablehead{
\colhead{} & \colhead{} &\colhead{EOS1} & \colhead{EOS2} & \colhead{EOS3}
}
\startdata
 Constituents & Uniform matter & $n$, $p$, $\alpha$      & $n$, $p$, $\alpha$      & $n$, $p$, $\alpha$, $\Lambda$ \\
          & Non-uniform matter & $n$, $p$, $\alpha$, $A$ & $n$, $p$, $\alpha$, $A$ & $n$, $p$, $\alpha$, $A$ \\
\hline
 $T$ & Range            & $ -1.0 \leq \log_{10}(T) \leq 2.0 $
                        & $ -1.0 \leq \log_{10}(T) \leq 2.6 $
                        & $ -1.0 \leq \log_{10}(T) \leq 2.6 $ \\
(MeV) & Grid spacing    & $\Delta \log_{10}(T)\simeq 0.1$
                        & $\Delta \log_{10}(T)=0.04$
                        & $\Delta \log_{10}(T)=0.04$ \\
     & Points           & 32 (including $T=0$) & 92 (including $T=0$) & 92 (including $T=0$) \\
\hline
       & Range          & $ -2 \leq \log_{10}(Y_p) \leq -0.25 $
                        & $ 0 \leq Y_p \leq 0.65 $
                        & $ 0 \leq Y_p \leq 0.65 $   \\
 $Y_p$ & Grid spacing   & $\Delta \log_{10}(Y_p)=0.025$
                        & $\Delta Y_p=0.01$
                        & $\Delta Y_p=0.01$ \\
     & Points           & 72 (including $Y_p=0$) & 66 & 66 \\
\hline
 $\rho_B$ & Range       & $  5.1 \leq \log_{10}(\rho_B) \leq 15.4 $
                        & $  5.1 \leq \log_{10}(\rho_B) \leq 16 $
                        & $  5.1 \leq \log_{10}(\rho_B) \leq 16 $  \\
 $\rm{(g\,cm^{-3})}$ & Grid spacing & $\Delta \log_{10}(\rho_B)\simeq 0.1 $
                                & $\Delta \log_{10}(\rho_B) = 0.1 $
                                & $\Delta \log_{10}(\rho_B) = 0.1 $ \\
     & Points                   & 104 & 110 & 110 \\
\hline
   Reference  &         & \citet{shen98b} & This paper & This paper  \\
\enddata
\tablecomments{EOS1 has been described in~\citet{shen98b},
while EOS2 and EOS3 are given in this study.
EOS1 and EOS2 take into account only the nucleon degree of freedom,
while EOS3 includes additional contributions from $\Lambda$ hyperons.
The non-uniform matter is modeled as a mixture of
free neutrons ($n$), free protons ($p$), alpha-particles ($\alpha$),
and a single species of heavy nuclei ($A$).
The $\Lambda$ fraction in non-uniform matter is quite small,
so we neglect the contribution from $\Lambda$ hyperons in the
non-uniform matter phase of EOS3.
In addition, we add the results of zero temperature,
this gives a total of 92 points for $T$ in EOS2 and EOS3.}
\end{deluxetable}

%%%%%%%%%%%%%%
\begin{deluxetable}{cccccccccc}
\tabletypesize{\scriptsize}
\tablecaption{The parameter set TM1 for the RMF Lagrangian
\label{tab:2}}
\tablewidth{0cm}
\tablehead{
 \colhead{$M$} & \colhead{$m_\sigma$} & \colhead{$m_\omega$} & \colhead{$m_\rho$} &
 \colhead{$g_\sigma$} & \colhead{$g_\omega$} & \colhead{$g_\rho$} &
 \colhead{$g_2$ ($\rm{fm}^{-1}$)} & \colhead{$g_3$} & \colhead{$c_3$}
}
\startdata
 938.0 & 511.19777 & 783.0 & 770.0 & 10.02892 & 12.61394 & 4.63219 & -7.23247 & 0.61833 & 71.30747 \\
\enddata
\tablecomments{The masses are given in $\rm{MeV}$.}
\end{deluxetable}

%%%%%%%%%%%%%%
\begin{deluxetable}{cccccccccccccccccc}
\tabletypesize{\tiny}
\rotate
\tablecaption{The main EOS table of EOS2
\label{tab:3}}
\tablewidth{0cm}
\tablehead{
 \colhead{$\log_{10}(T)$} & \colhead{$T$} &
 \colhead{$\log_{10}(\rho_B)$} & \colhead{$n_B$} & \colhead{$Y_p$} &
 \colhead{$F$}     &  \colhead{$E_{\rm{int}}$}   & \colhead{$S$}   &
 \colhead{$A$}     &  \colhead{$Z$}    & \colhead{$M^{*}_N$}       &
 \colhead{$X_n$}   &  \colhead{$X_p$}  & \colhead{$X_{\alpha}$}    & \colhead{$X_A$} &
 \colhead{$p$}     &  \colhead{$\mu_n$}& \colhead{$\mu_p$} \\
 \colhead{(MeV)}   &  \colhead{(MeV)}  &
 \colhead{($\rm{g\,cm^{-3}}$)}     &  \colhead{($\rm{fm^{-3}}$)} & \colhead{}   &
 \colhead{(MeV)}   &  \colhead{(MeV)}  &  \colhead{($k_B$)}  &
 \colhead{}        &  \colhead{}       &  \colhead{(MeV)}    &
 \colhead{}        &  \colhead{}       &  \colhead{}         &  \colhead{} &
 \colhead{($\rm{MeV\,fm^{-3}}$)} &  \colhead{(MeV)} &  \colhead{(MeV)}
}
\startdata
-1.0 & 0.1 &  5.1 & 7.581E-11 & 0.01 & -1.524 & 6.408 & 14.27 & 91.37 &  29.40 & 938 & 0.9689 & 0 & 0 & 0.03107 & 7.344E-12 & -1.222 & -21.73 \\
-1.0 & 0.1 &  5.2 & 9.544E-11 & 0.01 & -1.502 & 6.408 & 14.04 & 91.39 &  29.32 & 938 & 0.9688 & 0 & 0 & 0.03116 & 9.245E-12 & -1.199 & -21.78 \\
-1.0 & 0.1 &  5.3 & 1.201E-10 & 0.01 & -1.480 & 6.408 & 13.82 & 91.56 &  29.33 & 938 & 0.9687 & 0 & 0 & 0.03121 & 1.164E-11 & -1.176 & -21.83
\enddata
\tablecomments{
It covers the temperature range $T=10^{-1}$--$10^{2.6}$ MeV
with the logarithmic grid spacing $\Delta \log_{10}(T/\rm{[MeV]})=0.04$
(91 points), the proton fraction range $Y_p=0.01$--$0.65$
with the linear grid spacing $\Delta Y_p = 0.01$ (65 points),
and the density range $\rho_B=10^{5.1}$--$10^{16}\,\rm{g\,cm^{-3}}$
with the logarithmic grid spacing
$\Delta \log_{10}(\rho_B/\rm{[g\,cm^{-3}]}) = 0.1$ (110 points).\\
(This table is available in its entirety in a machine-readable form in the online journal.
A portion with less digits is shown here for guidance regarding its form and content.)
}
\end{deluxetable}

%%%%%%%%%%%%%%
\begin{deluxetable}{cccccccccccccccccc}
\tabletypesize{\tiny}
\rotate
\tablecaption{The table of EOS2 at $T=0$
\label{tab:4}}
\tablewidth{0cm}
\tablehead{
 \colhead{$\log_{10}(T)$} & \colhead{$T$} &
 \colhead{$\log_{10}(\rho_B)$} & \colhead{$n_B$} & \colhead{$Y_p$} &
 \colhead{$F$}     &  \colhead{$E_{\rm{int}}$}   & \colhead{$S$}   &
 \colhead{$A$}     &  \colhead{$Z$}    & \colhead{$M^{*}_N$}       &
 \colhead{$X_n$}   &  \colhead{$X_p$}  & \colhead{$X_{\alpha}$}    & \colhead{$X_A$} &
 \colhead{$p$}     &  \colhead{$\mu_n$}& \colhead{$\mu_p$} \\
 \colhead{(MeV)}   &  \colhead{(MeV)}  &
 \colhead{($\rm{g\,cm^{-3}}$)}     &  \colhead{($\rm{fm^{-3}}$)} & \colhead{}   &
 \colhead{(MeV)}   &  \colhead{(MeV)}  &  \colhead{($k_B$)}  &
 \colhead{}        &  \colhead{}       &  \colhead{(MeV)}    &
 \colhead{}        &  \colhead{}       &  \colhead{}         &  \colhead{} &
 \colhead{($\rm{MeV\,fm^{-3}}$)} &  \colhead{(MeV)} &  \colhead{(MeV)}
}
\startdata
-100 & 0 &  5.1 & 7.581E-11 & 0.01 & -0.2446& 6.261 & 0 & 96.11 &  28.40 & 938 & 0.9661 & 0 & 0 & 0.03383 & 5.848E-16 & 3.571E-05 & -24.46 \\
-100 & 0 &  5.2 & 9.544E-11 & 0.01 & -0.2446& 6.261 & 0 & 96.11 &  28.40 & 938 & 0.9661 & 0 & 0 & 0.03383 & 8.889E-16 & 4.156E-05 & -24.46 \\
-100 & 0 &  5.3 & 1.201E-10 & 0.01 & -0.2446& 6.261 & 0 & 96.12 &  28.41 & 938 & 0.9661 & 0 & 0 & 0.03383 & 1.328E-15 & 4.770E-05 & -24.46
\enddata
\tablecomments{
It covers the proton fraction range $Y_p=0.01$--$0.65$
with the linear grid spacing $\Delta Y_p = 0.01$ (65 points),
and the density range $\rho_B=10^{5.1}$--$10^{16}\,\rm{g\,cm^{-3}}$
with the logarithmic grid spacing
$\Delta \log_{10}(\rho_B/\rm{[g\,cm^{-3}]}) = 0.1$ (110 points).\\
(This table is available in its entirety in a machine-readable form in the online journal.
A portion with less digits is shown here for guidance regarding its form and content.)
}
\end{deluxetable}

%%%%%%%%%%%%%%
\begin{deluxetable}{cccccccccccccccccc}
\tabletypesize{\tiny}
\rotate
\tablecaption{The table of EOS2 for $Y_p=0$
\label{tab:5}}
\tablewidth{0cm}
\tablehead{
 \colhead{$\log_{10}(T)$} & \colhead{$T$} &
 \colhead{$\log_{10}(\rho_B)$} & \colhead{$n_B$} & \colhead{$Y_p$} &
 \colhead{$F$}     &  \colhead{$E_{\rm{int}}$}   & \colhead{$S$}   &
 \colhead{$A$}     &  \colhead{$Z$}    & \colhead{$M^{*}_N$}       &
 \colhead{$X_n$}   &  \colhead{$X_p$}  & \colhead{$X_{\alpha}$}    & \colhead{$X_A$} &
 \colhead{$p$}     &  \colhead{$\mu_n$}& \colhead{$\mu_p$} \\
 \colhead{(MeV)}   &  \colhead{(MeV)}  &
 \colhead{($\rm{g\,cm^{-3}}$)}     &  \colhead{($\rm{fm^{-3}}$)} & \colhead{}   &
 \colhead{(MeV)}   &  \colhead{(MeV)}  &  \colhead{($k_B$)}  &
 \colhead{}        &  \colhead{}       &  \colhead{(MeV)}    &
 \colhead{}        &  \colhead{}       &  \colhead{}         &  \colhead{} &
 \colhead{($\rm{MeV\,fm^{-3}}$)} &  \colhead{(MeV)} &  \colhead{(MeV)}
}
\startdata
-100 & 0 &  5.1 & 7.581E-11 & 0 & 2.281E-03& 6.508 & 0 & 0 &  0 & 938 & 1.0 & 0 & 0 & 0 & -1.703E-13 & 3.553E-05 & -938 \\
-100 & 0 &  5.2 & 9.544E-11 & 0 & 2.240E-03& 6.508 & 0 & 0 &  0 & 938 & 1.0 & 0 & 0 & 0 & -2.099E-13 & 4.142E-05 & -938 \\
-100 & 0 &  5.3 & 1.201E-10 & 0 & 1.151E-03& 6.507 & 0 & 0 &  0 & 938 & 1.0 & 0 & 0 & 0 & -1.323E-13 & 4.828E-05 & -938
\enddata
\tablecomments{
It covers the temperature range $T=10^{-1}$--$10^{2.6}$ MeV
with the logarithmic grid spacing $\Delta \log_{10}(T/\rm{[MeV]})=0.04$
(92 points including $T=0$) and the density range $\rho_B=10^{5.1}$--$10^{16}\,\rm{g\,cm^{-3}}$
with the logarithmic grid spacing
$\Delta \log_{10}(\rho_B/\rm{[g\,cm^{-3}]}) = 0.1$ (110 points).\\
(This table is available in its entirety in a machine-readable form in the online journal.
A portion with less digits is shown here for guidance regarding its form and content.)
}
\end{deluxetable}

%%%%%%%%%%%%%%
\begin{deluxetable}{cccccccccccccccccccc}
\tabletypesize{\tiny}
\rotate
\tablecaption{The main EOS table of EOS3,
which has the same ranges and grids for $T$, $Y_p$, and $\rho_B$ as the one of EOS2 given in Table 3
\label{tab:6}}
\tablewidth{0cm}
\tablehead{
 \colhead{$\log_{10}(T)$} & \colhead{$T$} &
 \colhead{$\log_{10}(\rho_B)$} & \colhead{$n_B$} & \colhead{$Y_p$} &
 \colhead{$F$}     &  \colhead{$E_{\rm{int}}$}   & \colhead{$S$}   &
 \colhead{$A$}     &  \colhead{$Z$}    & \colhead{$M^{*}_N$}       &
 \colhead{$X_n$}   &  \colhead{$X_p$}  & \colhead{$X_{\alpha}$}    & \colhead{$X_A$} &
 \colhead{$p$}     &  \colhead{$\mu_n$}& \colhead{$\mu_p$}         &
 \colhead{$M^{*}_{\Lambda}$}           & \colhead{$X_{\Lambda}$}
 \\
 \colhead{(MeV)}   &  \colhead{(MeV)}  &
 \colhead{($\rm{g\,cm^{-3}}$)}     &  \colhead{($\rm{fm^{-3}}$)} & \colhead{}   &
 \colhead{(MeV)}   &  \colhead{(MeV)}  &  \colhead{($k_B$)}  &
 \colhead{}        &  \colhead{}       &  \colhead{(MeV)}    &
 \colhead{}        &  \colhead{}       &  \colhead{}         &  \colhead{} &
 \colhead{($\rm{MeV\,fm^{-3}}$)} &  \colhead{(MeV)} &  \colhead{(MeV)} &
 \colhead{(MeV)}   &  \colhead{}
}
\startdata
-1.0 & 0.1 &  5.1 & 7.581E-11 & 0.01 & -1.524 & 6.408 & 14.27 & 91.37 &  29.40 & 938 & 0.9689 & 0 & 0 & 0.03107 & 7.344E-12 & -1.222 & -21.73 & 1115.7 & 0 \\
-1.0 & 0.1 &  5.2 & 9.544E-11 & 0.01 & -1.502 & 6.408 & 14.04 & 91.39 &  29.32 & 938 & 0.9688 & 0 & 0 & 0.03116 & 9.245E-12 & -1.199 & -21.78 & 1115.7 & 0 \\
-1.0 & 0.1 &  5.3 & 1.201E-10 & 0.01 & -1.480 & 6.408 & 13.82 & 91.56 &  29.33 & 938 & 0.9687 & 0 & 0 & 0.03121 & 1.164E-11 & -1.176 & -21.83 & 1115.7 & 0
\enddata
\tablecomments{
Comparing with EOS2, two additional columns are included,
which are the effective $\Lambda$ mass $M^{*}_{\Lambda}$ and the $\Lambda$ fraction $X_{\Lambda}$.\\
(This table is available in its entirety in a machine-readable form in the online journal.
A portion with less digits is shown here for guidance regarding its form and content.)
}
\end{deluxetable}

%%%%%%%%%%%%%%
\begin{deluxetable}{cccccccccccccccccccc}
\tabletypesize{\tiny}
\rotate
\tablecaption{The table of EOS3 at $T=0$,
which has the same ranges and grids for $Y_p$ and $\rho_B$ as the one of EOS2 given in Table 4
\label{tab:7}}
\tablewidth{0cm}
\tablehead{
 \colhead{$\log_{10}(T)$} & \colhead{$T$} &
 \colhead{$\log_{10}(\rho_B)$} & \colhead{$n_B$} & \colhead{$Y_p$} &
 \colhead{$F$}     &  \colhead{$E_{\rm{int}}$}   & \colhead{$S$}   &
 \colhead{$A$}     &  \colhead{$Z$}    & \colhead{$M^{*}_N$}       &
 \colhead{$X_n$}   &  \colhead{$X_p$}  & \colhead{$X_{\alpha}$}    & \colhead{$X_A$} &
 \colhead{$p$}     &  \colhead{$\mu_n$}& \colhead{$\mu_p$}         &
 \colhead{$M^{*}_{\Lambda}$}           & \colhead{$X_{\Lambda}$}
 \\
 \colhead{(MeV)}   &  \colhead{(MeV)}  &
 \colhead{($\rm{g\,cm^{-3}}$)}     &  \colhead{($\rm{fm^{-3}}$)} & \colhead{}   &
 \colhead{(MeV)}   &  \colhead{(MeV)}  &  \colhead{($k_B$)}  &
 \colhead{}        &  \colhead{}       &  \colhead{(MeV)}    &
 \colhead{}        &  \colhead{}       &  \colhead{}         &  \colhead{} &
 \colhead{($\rm{MeV\,fm^{-3}}$)} &  \colhead{(MeV)} &  \colhead{(MeV)} &
 \colhead{(MeV)}   &  \colhead{}
}
\startdata
-100 & 0 &  5.1 & 7.581E-11 & 0.01 & -0.2446& 6.261 & 0 & 96.11 &  28.40 & 938 & 0.9661 & 0 & 0 & 0.03383 & 5.848E-16 & 3.571E-05 & -24.46  & 1115.7 & 0  \\
-100 & 0 &  5.2 & 9.544E-11 & 0.01 & -0.2446& 6.261 & 0 & 96.11 &  28.40 & 938 & 0.9661 & 0 & 0 & 0.03383 & 8.889E-16 & 4.156E-05 & -24.46  & 1115.7 & 0 \\
-100 & 0 &  5.3 & 1.201E-10 & 0.01 & -0.2446& 6.261 & 0 & 96.12 &  28.41 & 938 & 0.9661 & 0 & 0 & 0.03383 & 1.328E-15 & 4.770E-05 & -24.46  & 1115.7 & 0
\enddata
\tablecomments{
(This table is available in its entirety in a machine-readable form in the online journal.
A portion with less digits is shown here for guidance regarding its form and content.)
}
\end{deluxetable}

%%%%%%%%%%%%%%
\begin{deluxetable}{cccccccccccccccccccc}
\tabletypesize{\tiny}
\rotate
\tablecaption{The table of EOS3 for $Y_p=0$,
which has the same ranges and grids for $T$ and $\rho_B$ as the one of EOS2 given in Table 5
\label{tab:8}}
\tablewidth{0cm}
\tablehead{
 \colhead{$\log_{10}(T)$} & \colhead{$T$} &
 \colhead{$\log_{10}(\rho_B)$} & \colhead{$n_B$} & \colhead{$Y_p$} &
 \colhead{$F$}     &  \colhead{$E_{\rm{int}}$}   & \colhead{$S$}   &
 \colhead{$A$}     &  \colhead{$Z$}    & \colhead{$M^{*}_N$}       &
 \colhead{$X_n$}   &  \colhead{$X_p$}  & \colhead{$X_{\alpha}$}    & \colhead{$X_A$} &
 \colhead{$p$}     &  \colhead{$\mu_n$}& \colhead{$\mu_p$}         &
 \colhead{$M^{*}_{\Lambda}$}           & \colhead{$X_{\Lambda}$}
 \\
 \colhead{(MeV)}   &  \colhead{(MeV)}  &
 \colhead{($\rm{g\,cm^{-3}}$)}     &  \colhead{($\rm{fm^{-3}}$)} & \colhead{}   &
 \colhead{(MeV)}   &  \colhead{(MeV)}  &  \colhead{($k_B$)}  &
 \colhead{}        &  \colhead{}       &  \colhead{(MeV)}    &
 \colhead{}        &  \colhead{}       &  \colhead{}         &  \colhead{} &
 \colhead{($\rm{MeV\,fm^{-3}}$)} &  \colhead{(MeV)} &  \colhead{(MeV)} &
 \colhead{(MeV)}   &  \colhead{}
}
\startdata
-100 & 0 &  5.1 & 7.581E-11 & 0 & 2.281E-03& 6.508 & 0 & 0 &  0 & 938 & 1.0 & 0 & 0 & 0 & -1.703E-13 & 3.553E-05 & -938  & 1115.7 & 0 \\
-100 & 0 &  5.2 & 9.544E-11 & 0 & 2.240E-03& 6.508 & 0 & 0 &  0 & 938 & 1.0 & 0 & 0 & 0 & -2.099E-13 & 4.142E-05 & -938  & 1115.7 & 0 \\
-100 & 0 &  5.3 & 1.201E-10 & 0 & 1.151E-03& 6.507 & 0 & 0 &  0 & 938 & 1.0 & 0 & 0 & 0 & -1.323E-13 & 4.828E-05 & -938  & 1115.7 & 0
\enddata
\tablecomments{
(This table is available in its entirety in a machine-readable form in the online journal.
A portion with less digits is shown here for guidance regarding its form and content.)
}
\end{deluxetable}

%%%%%%%%%%%%%%%%%%%%%%%%%%%%%%%%%%%%%%%%%%%%%%%%%%%%%%%%%%%%%%%%%%%%%%%%%%%%%%%%
\clearpage
%%%%%%%%%%%%%%

%%%%%%%%%%%%%%

%%%%%%%%%%%%%%

%%%%%%%%%%%%%%

%%%%%%%%%%%%%%

%%%%%%%%%%%%%%

%%%%%%%%%%%%%%

%%%%%%%%%%%%%%

%%%%%%%%%%%%%%

%%%%%%%%%%%%%%

%%%%%%%%%%%%%%%%%%%%%%%%%%%%%%%%%%%%%%%%%%%%%%%%%%%%%%%%%%%%%%%%%%%%%%%%%%%%%%%%
\end{document}